\documentclass[twocolumn]{aastex701}

\shorttitle{Ionized carbon streamers in protocluster core SPT2349$-$56}
\shortauthors{N. Sulzenauer et al.}

\begin{document}

\title{Bright [C\,{\sc ii}]158µm Streamers as a Beacon for Giant Galaxy Formation in SPT2349$\mathbf{-}$56 at $\mathbf{z=4.3}$}

\correspondingauthor{Nikolaus Sulzenauer}

\author[0000-0002-3187-1648]{Nikolaus Sulzenauer }
\altaffiliation{Member of the International Max Planck Research School (IMPRS) for Astronomy and Astrophysics at the Universities of Bonn and Cologne.}
\affiliation{Max-Planck-Institut für Radioastronomie, Auf dem Hügel 69, 53121 Bonn, Germany}
\email[show]{nsulzenauer@mpifr-bonn.mpg.de}

\author[0000-0003-4678-3939]{Axel Weiß}
\affiliation{Max-Planck-Institut für Radioastronomie, Auf dem Hügel 69, 53121 Bonn, Germany}
\email{aweiss@mpifr-bonn.mpg.de}

\author[0009-0008-8718-0644]{Ryley Hill}
\affiliation{Department of Physics and Astronomy, University of British Columbia, Vancouver, BC, V6T1Z1, Canada}
\email{ryleyhill@phas.ubc.ca}

\author[0000-0002-8487-3153]{Scott C. Chapman}
\affiliation{Department of Physics and Atmospheric Science, Dalhousie University, Halifax, NS, B3H 4R2, Canada}
\affiliation{NRC Herzberg Astronomy and Astrophysics, 5071 West Saanich Rd, Victoria, BC, V9E 2E7, Canada}
\affiliation{Department of Physics and Astronomy, University of British Columbia, Vancouver, BC, V6T1Z1, Canada}
\email{scott.chapman.dal@gmail.com}

\author[0000-0002-6290-3198]{Manuel Aravena}
\altaffiliation{Millenium Nucleus for Galaxies (MINGAL)}
\affiliation{Instituto de Estudios Astrof\'{\i}sicos, Facultad de Ingenier\'{\i}a y Ciencias, Universidad Diego Portales, Av. Ej\'ercito 441, Santiago, Chile}
\email{manuel.aravenaa@mail.udp.cl}

\author[0000-0002-9993-3796]{Veronica~J. Dike}
\affiliation{Department of Astronomy, University of Illinois, 1002 West Green St., Urbana, IL 61801, USA}
\email{vdike2@illinois.edu}

\author[0000-0002-0933-8601]{Anthony Gonzalez}
\affiliation{Department of Astronomy, University of Florida, 211 Bryant Space Science Center, Gainesville, FL 32611-2055, USA}
\email{anthonyhg@astro.ufl.edu}

\author[0009-0000-8306-1885]{Duncan MacIntyre}
\affiliation{Department of Physics and Astronomy, University of British Columbia, Vancouver, BC, V6T1Z1, Canada}
\affiliation{Department of Physics, University of Toronto, 60 St George Street, Toronto, ON, M5S1A7, Canada}
\email{dm1@student.ubc.ca}

\author[0000-0002-7064-4309]{Desika Narayanan}
\affiliation{Department of Astronomy, University of Florida, 211 Bryant Space Science Center, Gainesville, FL 32611-2055, USA}
\affiliation{Cosmic Dawn Center at the Niels Bohr Institute, University of Copenhagen and DTU-Space, Technical University of Denmark, Denmark}
\email{desika.narayanan@gmail.com}

\author[0000-0001-7946-557X]{Kedar~A. Phadke}
\affiliation{Department of Astronomy, University of Illinois, 1002 West Green St., Urbana, IL 61801, USA}
\affiliation{Center for AstroPhysical Surveys, National Center for Supercomputing Applications, 1205 West Clark Street, Urbana, IL 61801, USA}
\affiliation{NSF-Simons AI Institute for the Sky (SkAI), 172 E. Chestnut St., Chicago, IL 60611, USA}
\email{kphadke2@illinois.edu}

\author[0009-0003-6250-1396]{Vismaya~R. Pillai}
\affiliation{Department of Physics and Astronomy, University of British Columbia, Vancouver, BC, V6T1Z1, Canada}
\email{vpillai@student.ubc.ca}

\author[0000-0001-8598-064X]{Ana C. Posses}
\affiliation{Department of Physics and Astronomy and George P. and Cynthia Woods Mitchell Institute for Fundamental Physics and Astronomy, Texas A\&M University, 4242 TAMU, College Station, TX 77843-4242, USA}
\email{ana.posses@mail.udp.cl}

\author[0000-0002-1619-8555]{Douglas Rennehan}
\affiliation{Center for Computational Astrophysics, Flatiron Institute, 162 Fifth Avenue, New York, NY 10010, USA}
\email{douglas.rennehan@gmail.com}

\author[0000-0003-4357-3450]{Amélie Saintonge}
\affiliation{Max-Planck-Institut für Radioastronomie, Auf dem Hügel 69, 53121 Bonn, Germany}
\affiliation{Department of Physics and Astronomy, University College London, Gower Street, London, WC1E 6BT, UK}
\email{asaintonge@mpifr-bonn.mpg.de}

\author[0000-0003-3256-5615]{Justin S. Spilker}
\affiliation{Department of Physics and Astronomy and George P. and Cynthia Woods Mitchell Institute for Fundamental Physics and Astronomy, Texas A\&M University, 4242 TAMU, College Station, TX 77843-4242, USA}
\email{jspilker@tamu.edu}

\author[0000-0001-6629-0379]{Manuel Solimano}
\affiliation{Instituto de Estudios Astrof\'{\i}sicos, Facultad de Ingenier\'{\i}a y Ciencias, Universidad Diego Portales, Av. Ej\'ercito 441, Santiago, Chile}
\email{cosmo.manus@gmail.com}

\author{Joel Tsuchitori}
\affiliation{Department of Physics and Astronomy, University of British Columbia, Vancouver, BC, V6T1Z1, Canada}
\email{tsuchitorijoel@gmail.com}

\author[0000-0001-7192-3871]{Joaquin D. Vieira}
\affiliation{Department of Astronomy, University of Illinois, 1002 West Green St., Urbana, IL 61801, USA}
\affiliation{Center for AstroPhysical Surveys, National Center for Supercomputing Applications, 1205 West Clark Street, Urbana, IL 61801, USA}
\email{jvieira@illinois.edu}

\author[0000-0001-7192-3871]{David Vizgan}
\affiliation{Department of Astronomy, University of Illinois, 1002 West Green St., Urbana, IL 61801, USA}
\email{dvizgan2@illinois.edu}

\author[0000-0002-6922-469X]{Dazhi Zhou}
\affiliation{Department of Physics and Astronomy, University of British Columbia, Vancouver, BC, V6T1Z1, Canada}
\email{dzhou@astro.ubc.ca}

\begin{abstract} 

Observations of extreme starbursts, often located in the cores of protoclusters, challenge the classical bottom-up galaxy formation paradigm. Giant elliptical galaxies at $z=0$ must have assembled rapidly, possibly within few~100\,Myr through an extreme growth phase at high-redshift, characterized by elevated star-formation rates of several thousand solar masses per year distributed over concurrent, gas-rich mergers. We present a novel view of the $z=4.3$ protocluster core SPT2349$-$56 from sensitive multi-cycle ALMA dust continuum and [C\,{\sc ii}]158$\mu$m line observations. Distributed across 60\,kpc, a highly structured gas reservoir with a line luminosity of $L_\mathrm{[CII]}=3.0\pm0.2\times10^9$\,$L_\sun$ and an inferred cold gas mass of $M_\mathrm{gas}= 8.9\pm0.7\times10^{9}$\,$M_\sun$ is found surrounding the central massive galaxy triplet. Like ``beads on a string'', the newly-discovered [C\,{\sc ii}] streamers fragment into a few kpc-spaced and turbulent clumps that have a similar column density as local Universe spiral galaxy arms at $\Sigma_\mathrm{gas}=20$--$60$\,$M_\sun\,\mathrm{pc}^{-2}$. For a dust temperature of 30\,K, the [C\,{\sc ii}] emission from the ejected clumps carry $\gtrsim$3\% of the FIR luminosity, translating into an exceptionally low mass-to-light ratio of $\alpha_\mathrm{[CII]}=2.95\pm0.3$\,$M_\sun$\,$L_\sun^{-1}$, indicative of shock-heated molecular gas. In phase space, about half of the galaxies in the protocluster core populate the same caustic as the [C\,{\sc ii}] streamers ($r/r_\mathrm{vir}\times|\Delta v|/\sigma_\mathrm{vir}\approx0.1$), suggesting angular momentum dissipation via tidal ejection while the brightest cluster galaxy (BCG) is assembling. Our findings provide new evidence for the importance of tidal ejections of [C\,{\sc ii}]-bright, shocked material following multiple major mergers that might represent a landmark phase in the $z\gtrsim4$ co-evolution of BCGs with their hot, metal enriched atmospheres.

\end{abstract}


\section{Introduction}\label{sec:intro}

Brightest cluster galaxies (BCGs) are located close to the bottom of the gravitational potential trough, embedded in a hot, X-ray-bright atmosphere, indicating co-evolution with the heating of the intracluster medium (ICM) \citep{Katayama2003,Dekel2006,Kravtsov2012}. However, it is still an open question how these massive galaxies formed in the distant past, with stellar ages pointing to a formation time before $z=4$ \citep{Ziegler1997, Collins2009,Willis2020, Kubo2021,Trudeau2021}, as their general properties considerably deviate from the tight scaling relations of less luminous cluster and field ellipticals \citep{Faber1976,Oegerle1991,Thomas2005,Thomas2010,Kormendy2009}.

Both gas-rich and dry mergers contribute to the formation of BCGs. In the earliest formation phases, extreme starbursts caused by gas-rich major mergers or a dissipative collapse of a cold gas reservoir are required \citep{Thomas2005,DeLucia2006,Naab2009}. In addition, dissipationless dry merging among cluster core galaxies may contribute to the formation of the outer stellar envelopes around BCGs and transforming them into cD galaxies \citep{DeLucia2006,DeLucia2007}.

Submillimeter galaxies (SMGs; \citealt{Blain2004,Chapman2005,Casey2013}) are the most vigorously star-forming, highly dust obscured galaxies, forming the extreme tail end of the distribution of dusty star-forming galaxies (DSFGs; \citealt{Casey2013}). With flux densities of $S_{850\mu\mathrm{m}}>5$\,mJy, SMGs are the likely progenitors of these massive cluster galaxies \citep{Weiss2009,Hickox2012,Toft2014,Garcia-Vergara2020}. Spatially coherent groups of DSFGs with synchronized star-formation histories were discovered in the distant Universe \citep{Casey2016}. These ``protocluster cores'' \citep{Overzier2016,Alberts2022} show the expected properties of proto-BCGs at $3\lesssim z\lesssim7$ \citep{Wang2021}, because they contain massive coeval galaxies within a few 100\,kiloparsec in projection. Particularly prevalent are hyperluminous infrared galaxies (HyLIRGs) with $L_\mathrm{IR}\geq10^{13}\,L_\sun$ 
indicating extreme star-formation rates of $\mathrm{SFR}\gtrsim1000\,M_\sun$\,yr$^{-1}$ \citep{Kennicutt1998} capable of assembling $10^{11-12}\,M_\sun$ in stellar mass within a few~100\,Myr \citep{Ivison2013,Oteo2018,Miller2018,Wang2025}. Owing to their rarity, just a small number of vigorously star-forming groups at $z\gtrsim3$ have been robustly identified on the sky \citep[e.g.][]{Ivison2013,Oteo2018,Miller2018,Jin2021}.

Alongside the cold gas accretion from the larger protocluster structures \citep{Dekel2006,Dekel2006b}, the enhanced gas-rich merging activity plays an important role in triggering and later quenching the coeval starbursts among galaxies separated by $\sim$1\,cMpc \citep{Casey2016}. Gas-rich major mergers are typically accompanied by the formation of tens of kpc-long tidal arms, resembling ``antennae'', tails, or shells \citep{Sanders1996} that are stripped during the first and second pericenter passage \citep{Toomre1972,Duc2008}. Tidal debris extending far into the circumgalactic medium (CGM) leads to a disturbed appearance in stellar and cold gas components \citep{Elmegreen2007,Teyssier2010,deBlok2018,Spilker2022}. 

Multiple mergers also foster supermassive black hole growth \citep{Hopkins2008,Diaz-Santos2018,Decarli2019,Wylezalek2022} and litter their halos with numerous metal rich tidal streamers \citep[see][]{Li2007,Bekki2001}. Such studies demonstrate that -- apart from AGN and star-formation feedback-driven winds \citep{Fujimoto2019,Herrera-Camus2021,Solimano2024} -- collision-induced kinetic shock-heating \citep{Appleton2013} and catastrophic tidal ejection of gas facilitate heating of the CGM and successively form the proto-ICM in protoclusters. Direct evidence for tidal ejection of gas from proto-BCGs is provided in observations of coherent gas streams between SMGs, reported in \citet{Emonts2013}, \citet{Ginolfi2020}, and \citet{Umehata2021}. Despite observational challenges, tidal ejections are among the main feedback mechanisms affecting halo mass growth by regulating the consumption of the molecular gas reservoir, star-formation activity, and redistributing shock heated gas \citep{Li2007,Teyssier2010,Webb2017,Puglisi2021,Spilker2022}.

DSFGs at high-$z$ can be studied in detail with the [C\,{\sc ii}]158$\mu$m ${^2}P_{3/2}\to{^2}P_{1/2}$ fine structure line transition at rest-frequency $1900.537$\,GHz of ionized carbon \citep{Carilli2013,Hodge2020}. 
Owing to the low ionizing potential of C$^{+}$ ($11.3$\,eV), low excitation energy ($E/k_\mathrm{B}\approx92$\,K), high elemental carbon abundance, and moderate optical depth, [C\,{\sc ii}]158$\mu$m is the prime cooling line of the cold interstellar medium at solar metallicity \citep[e.g.][]{Hollenbach1991}. This line carries (10$^3-$10$^4$)$\times$ more luminosity than CO($J\,{=}\,1\,{-}\,0$) and typically contributes $\lesssim$1\% of the dust blackbody FIR-cooling power \citep{Brauher2008,Stacey2010,Madden2020}. Thus, [C\,{\sc ii}] is routinely utilized at $z\gtrsim2$ to estimate molecular gas mass and/or SFRs \citep{Zanella2018,Bethermin2020,Schaerer2020,Dessauges-Zavadsky2020,Aravena2024}, despite important observational caveats \citep{Gullberg2015,Vizgan2022}. A dominant fraction of [C\,{\sc ii}] emission originates from the surface of far-ultraviolet-irradiated molecular clouds, so-called photo-dissociation regions (PDRs); while the rest is mainly produced in the hot ionized medium (HIM) \citep{Madden1997,Stacey2010,Gullberg2015,Accurso2017,Madden2020}.

Given that protocluster cores are rare, large survey volumes are required for their selection. One of the richest samples of high-$z$ SMGs is from the $\sim2500$\,deg$^2$ SPT-SZ survey \citep{Carlstrom2011,Everett2020}, comprising of a millimetric, multi-band selection of highly star-forming systems within a redshift range of $1.5<z<6.9$ \citep{Vieira2013, Weiss2013, Reuter2022}. About 10\% are not strongly lensed systems, but are resolved with APEX/LABOCA ($\lambda_\mathrm{obs}=870$\,$\mu$m) and ALMA at $<10\arcsec$-resolution into multiple continuum components \citep{Miller2018,Wang2021}. The resulting sample of protocluster core candidates (Hill et al. in prep.) contains the intrinsically 1.4\,mm-brightest source with $S_{1.4}=23.3$\,mJy \citep{Everett2020}, identified as \object{SPT-S~J234942-5638.2} or SPT2349$-$56 in short. It is composed of a northern ($S_{870}=22\pm1.4$\,mJy) and southern ($S_{870}=77\pm2.9$\,mJy) sub-group, separated by 350\,kpc \citep{Strandet2016}. \citet{Miller2018} discovered 15 ULIRGs in the system that were later supplemented in \citet{Hill2020} by nine additional [C\,{\sc ii}] line emitters at $z=4.303$. At an age of 1.4\,Gyr after the Big Bang, SPT2349$-$56 hosts a SFR of $\sim$6700\,$M_\sun$\,yr$^{-1}$ within 400\,kpc -- a record for a $\sim10^{13}$\,M$_\sun$ group sized halo. However, only $\sim$55\% of the APEX/LABOCA flux density could be recovered with ALMA \citep{Hill2020}, highlighting the important role of the multiphase circumgalactic gas reservoir in this protocluster core \citep[see also][]{Zhou2025}.

Non-cosmological hydro-simulations of the system showed that the extreme space density of SMGs is dynamically unstable and will rapidly collapse to form a single $M_\star\approx 10^{12}\,M_\sun$ proto-BCG within the next 100--300 Myr corresponding to $z=3$ \citep{Rennehan2020}. The galaxies are characterized by disturbed molecular disks \citep{Venkateshwaran2024} and show systematically warm ISM conditions \citep{Hughes2024}. 
Therefore, SPT2349$-$56 is a prime candidate to study \textit{in situ} massive galaxy formation. 

Few high-$z$ galaxy groups are known that display a comparably high rate of gravitational interactions \citep{Oteo2016,Diaz-Santos2018, Decarli2019, Umehata2021, Long2020}, but none with a similarly high level of coeval star-formation activity of $4\times10^4\,M_\sun$\,yr$^{-1}$\,Mpc$^{-3}$ \citep{Hill2020}. \citet{Chapman2024} found radio emission co-spatial with one of the central SMGs, evidence for a radio-loud AGN at the center of the assembling cluster, to which \citet{Vito2024} provided empirical support from X-ray observations. 
Characterizing rare objects such as SPT2349$-$56 can provide important constraints on high-$z$ massive galaxy formation models in the $n(M_\star>10^{11}\,M_\sun)=10^{-5}$\,cMpc$^{-3}$ regime, where many simulations fail to reproduce the observed number counts (\citealt{Lim2021,Schaye2023}; but also see \citealt{Remus2023,Rennehan2024,Kimmig2025}).

In this work, we build upon the serendipitous discovery of a giant [C\,{\sc ii}] arc found close to three ULIRGs in the core of SPT2349$-$56 as previously reported in \citet{Hill2020}. 
Hereafter, we adopt a flat $\Lambda$CDM cosmology from \citet{Planck2016} with parameters: $H_0 = 67.8$ km s$^{-1}$ Mpc$^{-1}$, $\Omega_m = 0.308$, and $\Omega_\Lambda = 1 - \Omega_m = 0.692$, and we use the \citet{Chabrier2003} initial mass function. At $z=4.303$, the angular diameter distance is 6.909\,kpc per arcsecond.

\section{Observations and data reduction}\label{sec:observations}

\subsection{ALMA observations}

Following-up on the initial single-pointing ALMA Band-7 Cycle-4 observing program (ID:  2016.1.00236.T; PI: S. Chapman), published in \citet{Miller2018}, we observed a deeper, regularly-spaced, three pointing mosaic, covering the southern LABOCA core of SPT2349$-$56 in Cycle-5 (ID:2017.1.00273.S, PI S. Chapman), and at higher resolution with a six-point mosaic in Cycle-6 \citep{Venkateshwaran2024}.

\begin{table}[h!]
\setlength{\tabcolsep}{2pt}
\scriptsize
\center
\tablenum{1}
\caption{ALMA Band-7 observations of SPT2349$-$56.
\label{tab:observations}}
\begin{tabular}{lcccccc}
\tableline
\tableline
Cycle    & Mosaic       & Beam size       & Mean RMS & Channel\,$\Delta V$ &  Baselines$^\mathrm{a}$  \\
         &              & [arcsec$^2$]       & [mJy bm$^{-1}$] & [km\,s$^{-1}$]           &  (Q5--Q80)[m]  \\
\tableline
4  & 1-point   & 0.52$\times$0.50    &    1.21 & 13.00     &  40--207     \\
5  & 3-point   & 0.59$\times$0.49    &     0.33 & 13.08     &  27--176   \\
6  & 6-point   & 0.24$\times$0.17      &     0.45 & 13.07     &  59--544   \\
8  & 7-point   & 0.48$\times$0.43  &      0.34 & 13.07     &  34--226     \\
\tableline
4,5,8  & ``patchy"  & 0.47$\times$0.44    & 0.17 & 13.07     &  27--226    \\
\tableline
\end{tabular}
\tablecomments{$^\mathrm{a}$5$^\mathrm{th}$--80$^\mathrm{th}$ percentiles of baseline length distribution.}
\end{table}

Most recently, our ALMA Band-7 Cycle-8 program (ID:2021.1.01063.S, PI R. Hill) targeted [C\,{\sc ii}] line emission in an ultra-deep 10\,h integration with seven overlapping pointings arranged in a hexafoil pattern. The data was taken at an intermediate spatial resolution, in between the previous observations, and with the same spectral set-up. 
Observations were conducted between May 5$^\mathrm{th}$ to 14$^\mathrm{th}$ 2022 in excellent weather conditions of $\mathrm{pwv}=0.5$\,mm. 
See Tab.~\ref{tab:observations} for more details.

In this work, we combined previously obtained Band-7 $uv$ data from Cycle-4, Cycle-5, and Cycle-8 of the southern LABOCA source of SPT2349$-$56 into a deep [C\,{\sc ii}] spectral cube. Because our goal was to recover low-surface brightness emission, and in order to avoid over-resolving extended structures, the highest-resolution ALMA Cycle-6 data was not included in our combined [C\,{\sc ii}] data cube. From the 10 Cycle-8 pointings made of the outskirts of SPT2349$-$56, we included the four that were closest to the core to still provide useful information on the central galaxies. The primary beam illumination of the combined mosaic is shown in Fig.~\ref{fig:map}. 

Calibration was done by running the scripts provided by the observatory in the appropriate {\tt CASA}\footnote{\url{https://casa.nrao.edu}} version, and imaging was performed with {\tt CASA} using \texttt{tclean} with Briggs weighting with parameter $\mathtt{robust}=0.5$, resulting in a beam size of 0.47$\arcsec$$\times$0.44$\arcsec$ (3.3$\times$3.0\,kpc$^2$) at full width half maximum ($\mathrm{FWHM}$). During the cleaning, auto-multithresh masking was used, and cleaning was done down to 0.2\,mJy\,beam$^{-1}$, or about 1$\sigma$ in a single channel. Specifically, a frequency dependent RMS of 0.145--0.34\,mJy per channel per beam is obtained in the USB.
The maximum recoverable scale (MRS) of $\sim 6\arcsec$ or 41\,kpc at $z=4.3$ limits our sensitivity towards large-scale image features.

\subsection{Auxiliary data}\label{sec:auxi}

An ALMA [NII]205$\mu$m line cube was obtained in Cycle-5 observations (PI:Chapman, ID:2016.1.00236.T). Optical and NIR data sets as already discussed in \citet{Rotermund2021} and \citet{Hill2022} utilize Gemini/GMOS (project ID GS-2017B-Q-7, PI S. Chapman), {\it HST}/WFC3 (project ID 15701, PI S. Chapman), and {\it Spitzer}/IRAC (project IDs 13224 and 14216, PI S. Chapman). VLT/MUSE spectral imaging, targeting redshifted Ly$\alpha$ line emission is described in \citet{Apostolovski2022} (IDs 0100.A-0437(A) and 0100.A-0437(B), PI M. Aravena). Furthermore, APEX/LABOCA 870$\mu$m continuum maps were observed under project ID:299.A-5045(A), PI S. Chapman.

\subsection{Source selection}\label{sec:reduction}

\begin{figure*}[ht!]
\centering
\epsscale{1.0}
\includegraphics[width=1\textwidth]{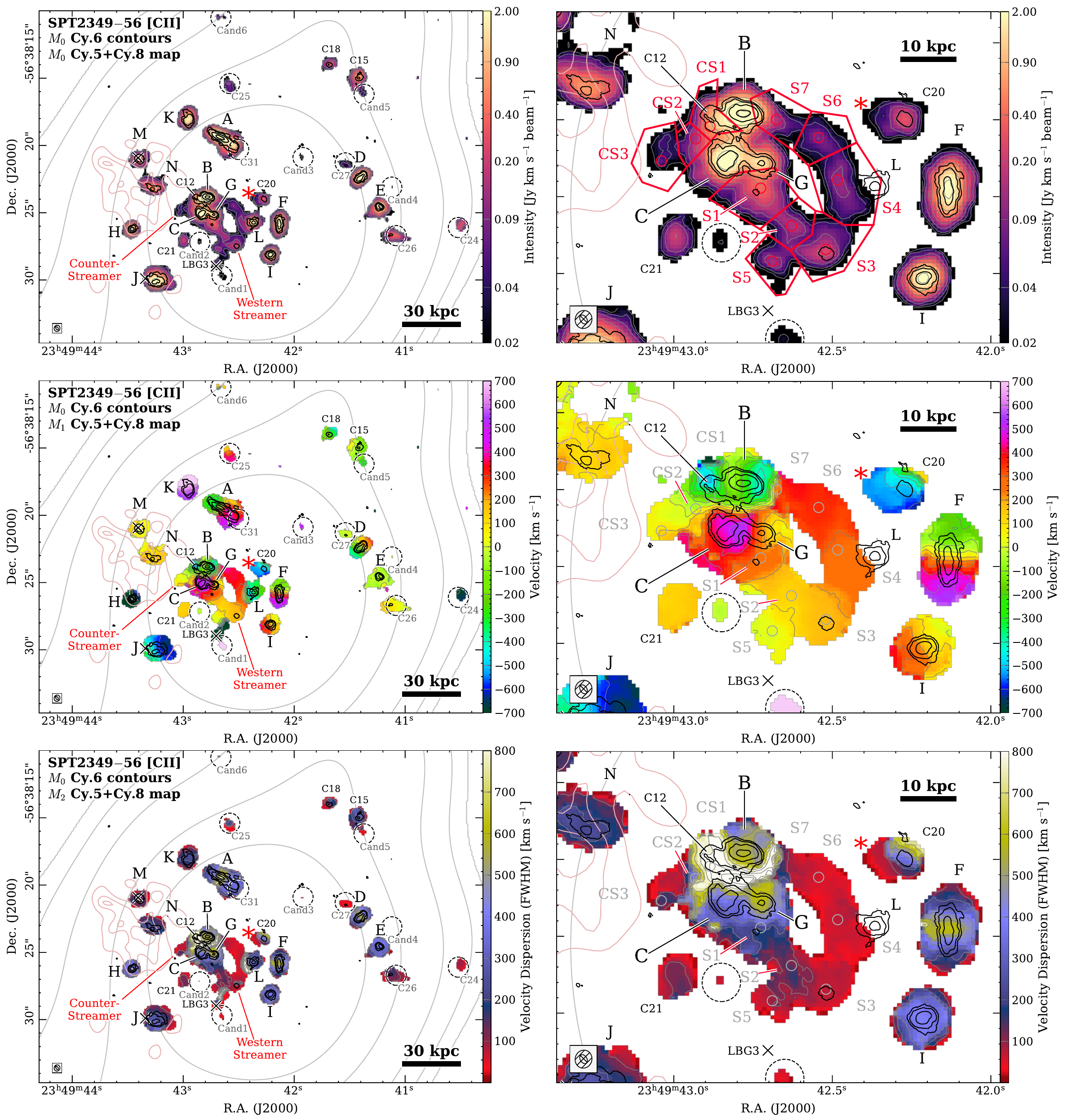}
\caption{Moment maps ($M_0$, $M_1$, and $M_2$ -- top to bottom row) of the protocluster core SPT2349$-$56 in [C\,{\sc ii}]158$\mu$m line emission. A zoomed-in version of the maps in the left column (230$\times$175~kpc$^2$) is shown on the right (80$\times$60~kpc$^2$). Black contours trace Cycle-6 [C\,{\sc ii}] emission at $\sim$0.25$\arcsec$ angular resolution with [0.03, 0.15, 0.6, 1.8, 3.6]\,Jy\,km\,s$^{-1}$\,beam$^{-1}$ intensity. Right column shows zoom-in on the central ULIRG triplet with labeled streamer segments (red polygons) and $M_0$ flux peaks (red circles). The 850$\mu$m-weighted center of the system \citep{Hill2020} is indicated by a the red asterisk. The ``Lyman-$\alpha$ blob'' \citep{Apostolovski2022} is marked in pink contours with [3, 5, 7]$\sigma$. New [C\,{\sc ii}] galaxy candidates are marked as dashed circles.  Central positions for LBGs \citep{Rotermund2021} are shown as crosses. SMGs `L' and `LBG3' are removed from the zoom-in map for clarity. Gray contours denote the primary beam at [$1/1.5$, $1/2$, $1/3$, $1/4$, $1/5$]$\times$ illumination. Note that due to the primary beam illumination pattern, the dilated masking procedure is $>$2$\times$ more restrictive for a source outside of the $1/2$ primary beam contour --- potentially masking low-surface brightness emission.\label{fig:map}}
\end{figure*}

Spectral moment-0 ($M_0$), moment-1 ($M_1$), and moment-2 ($M_2$) maps, shown in Fig.~\ref{fig:map} and Fig.~\ref{fig:momentszoom}, corresponding to spatially resolved velocity integrated line intensity ($S_\mathrm{[CII]}\Delta V$), velocity offset ($\Delta v_\mathrm{[CII]}$), and velocity dispersion ($\sigma_{v,\mathrm{[CII]}}$), respectively, were created from the combined ALMA [C\,{\sc ii}] cube over a velocity range of [$-$914,$+$1178]\,km\,s$^{-1}$. For comparison, the same procedure is conducted for the ALMA Cycle-6 data. All steps leading to the creation of the final cubes and of the moment maps was performed with the software package {\tt MIRIAD}\footnote{\url{https://www.atnf.csiro.au/computing/software/miriad/}} \citep{Sault1995} employing the following programmatic reduction steps. 

After dust continuum model subtraction (see Sec.~\ref{sec:continuum}), voxels are masked by a predefined $S/N$-cutoff to prevent contamination by false-positive sources due the frequency-dependent background noise in ALMA Band-7 (as a result of atmospheric O$_3$ line contamination) above a certain significance level. However, typical $\sigma$-clipping techniques -- employed at a 3--5$\sigma$ confidence level -- are notorious for filtering out extended, low-surface brightness features. 

To preserve diffuse emission features, a dilated masking technique similar to \citet{Rosolowsky2006} or \citet{Solimano2024} was employed. To bring the sources onto a homogeneous background intensity level, it was necessary to compensate for a significant \texttt{CLEAN} hole signal \citep{Pety2013} of $\int_{0}^{\Delta V} S^\mathrm{peak}_\mathrm{hole}\mathrm{d}v\approx-0.5\langle\sigma_\mathrm{[CII]}\rangle\Delta V$ within few hundred km\,s$^{-1}$ of $\Delta V$ in the combined cube. 
To avoid a frequency-dependent, inhomogeneous background level, a $\sigma=2$ mask is first created to limit intensity contributions from bright sources. Then, the \texttt{CLEAN} hole is corrected by least-squares minimization of a two-dimensional higher-order polynomial function, fitted independently per spectral channel to the masked intensity distribution. The subtracted cube is then cloned and clipped to a low and high-$\sigma$ level, respectively, and masked voxels are required to appear in two consecutive channels. Further, the high-$\sigma$ cube is ``dilated'' by convolution with the clean beam. Dilated cloned cubes are then combined by removing all voxel below the pre-defined $\sigma$-level. 

The {\tt MIRIAD} scripts for creating the analysis cubes and moment maps, as part of the software package \texttt{Clipper}, are publicly available as a {\tt GitHub} repository\footnote{\url{https://github.com/NiSZR/clipper}}. Voxels down to an equivalent of 2.3$\sigma$\,beam$^{-1}$\,channel$^{-1}$ or a sources with a specific line intensity at phase center $I_\mathrm{[CII]}\geq0.345$\,mJy\,beam$^{-1}$\,channel$^{-1}$ ($\nu=358$~GHz) in two consecutive channels ($\geq3.3$$\sigma$\,beam$^{-1}$ over 26\,km\,s$^{-1}$), are recovered.

\subsection{Dust continuum measurements}\label{sec:continuum}

A continuum map is created with \texttt{tclean} by collapsing all channels without line emission along the frequency axis of the lower side band (LSB) employing multifrequency synthesis. Many of the known [C\,{\sc ii}] sources show up as bright, marginally extended continuum sources. 
Given the large number of individually bright, extended SMGs, continuum sources are modeled in the image plane. We use two-dimensional Gaussian models from the {\tt MIRIAD} utility function {\tt imfit} to represent sources. 

The background RMS of the lower side band (LSB) continuum image, at central frequency $\nu=346.41$ GHz, is calculated over 47458 pixels ($\sim$$2030\times\Omega_\mathrm{bm}$) masking residual emission from the SMGs. 
When including the ultra-deep Cycle-8 data, the RMS sensitivity limit decreases by $\sim$66\% to $\sigma_{\nu}=13.5$\,$\mu$Jy\,beam$^{-1}$ compared to the continuum map presented in \citet{Hill2020}. Flux densities measurements $S_\nu$ are corrected by dividing with the primary beam response $f_\mathrm{pb}\in(0.2,1]$, shown in Fig.~\ref{fig:continuum}.

A simultaneous six-component two dimensional Gaussian model fit in the image plane is performed with \texttt{MIRIAD}, representing the `B'-`C'-`G' system, and galaxy `C12' \citep{Miller2018,Hill2020}. Peak flux density, center position, position angle, and semi major/minor axis values are left unconstrained. For the remaining sources, point-like models are employed to fit a bridge-like component between `C' and `G', and source `C16' (henceforth called 'S1') where we see excess residuum in the continuum map (see Fig.~\ref{fig:continuum}). Here, the parameterized beam with $\theta_\mathrm{mi}=0.44\arcsec$, $\theta_\mathrm{mj}=0.47\arcsec$, and $\mathrm{PA}=39.6\degr$ is scaled to the source amplitudes with unconstrained central positions.

To estimate star-formation rates (SFR; see Tab.~\ref{tab:results_continuum}) from the far-infrared (FIR) luminosity, we integrate \begin{eqnarray}\label{equ:fir} L_\mathrm{FIR}\equiv L_{\mathrm{42-500}\mu m}&=4\pi D_{L}^2(z)\int_{\nu_2}^{\nu_1}S_\nu \mathrm{d}\nu\end{eqnarray} within $\lambda_{1}=42$\,$\mu$m to $\lambda_{2}=500$\,$\mu$m using $\nu_i=c/\lambda_i$ and by fitting a spectral energy distribution (SED) model $S_\nu$ to individual dust measurements. For consistency with previous studies, a conversion factor of $0.95\times10^{-10}$ $M_\sun$\,yr$^{-1}$\,$L_\sun^{-1}$ \citep{Kennicutt1998} for a \citet{Chabrier2003} IMF is employed. Currently, we are incapable of robustly measuring FIR luminosities for individual SMGs since no spatially resolved data at wavelengths shorter than the peak of the dust SED were obtained so far. Dust masses and a weighted average temperature of $T_\mathrm{d}=39.6$\,K was estimated from resolved $S_{850\mu\mathrm{m}}/S_{3.2\mathrm{mm}}$ colors \citep{Hill2020}.

\subsection{[CII] line measurements}\label{sec:measure_lines}

\begin{figure}[ht!]
\flushright
\epsscale{1.0}
\includegraphics[width=0.475\textwidth]{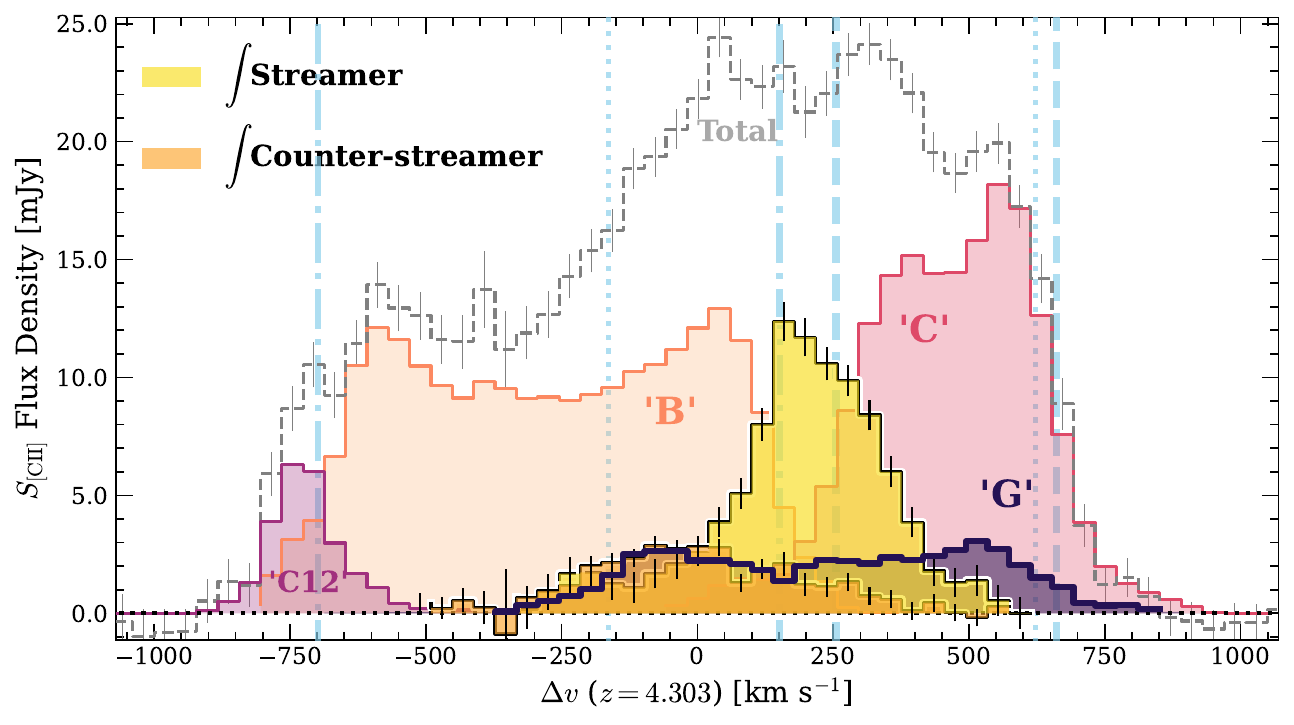}\vspace{-1.5mm}
\includegraphics[width=0.48\textwidth]{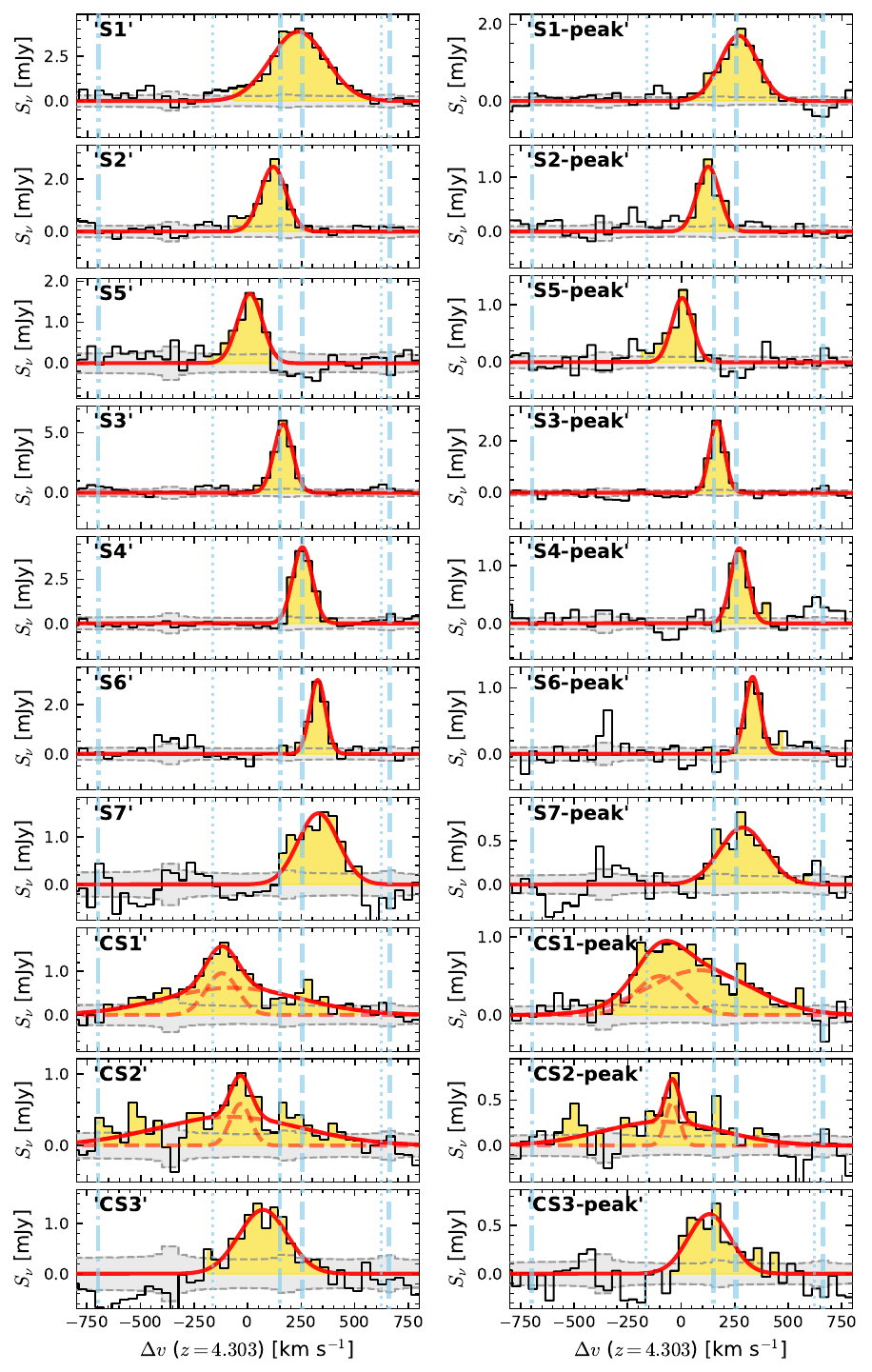}
\caption{Spatially integrated [C\,{\sc ii}] spectra of the western streamer (yellow) and counter-streamer (orange). The central galaxy triplet `B'-`C'-`G', including 'C12', is shown (top panel) and extracted spectra from the segments (left column) and peak positions (right column). Solid (dashed) red curves are Gaussian (double-Gaussian) model best fits to the observed data. The vertical blue lines correspond to the velocity domains enclosing $\sim$95 per cent of the flux of galaxies 'B', 'C', and 'G'. The RMS noise per binned channels is shown as grey shades.   \label{fig:spectra}}
\end{figure}

Polygonal shapes are manually defined on the moment-0 map, as shown in Fig.~\ref{fig:map}. The [C\,{\sc ii}] line flux density is then extracted using {\tt CARTA}\footnote{\url{https://cartavis.org/}} (version 2.0; \citealt{Comrie2021}) in units of Jansky per pixels contained within the area of the polygonal segments $\Omega_\mathrm{seg}$, i.e. $S_\mathrm{[CII]}=\int_{\Omega_\mathrm{seg}} I_\mathrm{[CII]}$. Line flux density contribution from nearby SMGs are modeled and subtracted out by using two-dimensional Gaussian component fitting within the \texttt{MIRIAD} software package. \texttt{Clipper} produces a spectral noise cube with the uncertainty of the [C\,{\sc ii}] line emission as a function of frequency. Factoring in $f_\mathrm{pb}$, extracted spectra and corresponding noise contributions are shown in Fig.~\ref{fig:spectra}. 

To improve data fidelity, the extracted signal and noise spectra are first binned by a factor of three, corresponding to a line-of-sight velocity width of $\delta v=3\times v_\mathrm{ch}\approx39.2$\,km\,s$^{-1}$. We then fitted each line with a Gaussian line profile using a least-squares minimization via the Levenberg-Marquardt algorithm through {\tt Python} package {\tt scipy.optimize}\footnote{The Levenberg-Marquardt least-squares algorithm is employed via the Python function \texttt{curve\_fit} from the method \texttt{scipy.optimize} (\url{https://docs.scipy.org/doc/scipy/reference/generated/scipy.optimize.least squares.html}).}. The velocity dispersion is converted to $\mathrm{FWHM}$, and corrected by the binned channel width via $\mathrm{FWHM}_\mathrm{[CII]}=\sqrt{8\ln{(2)}\sigma_v^2-\delta v^2}$. 
Integrating the Gaussian model fit yields the line intensity, i.e. $\int_{-\infty}^{+\infty} S_\mathrm{[CII]}^\mathrm{fit}\mathrm{d}v=S_\mathrm{[CII]}\Delta V$ in Jy\,km\,s$^{-1}$. This line fitting method has the advantage of directly including a weighted uncertainty on the extracted line flux density as an input for {\tt curve\_fit}. 
The line luminosity is estimated from the line intensity as 
\begin{eqnarray}\label{equ:lcii}
L_\mathrm{[CII]}&=&\frac{4\pi}{c}D_L^2(1+z)^{-1}\nu_\mathrm{[CII]}S_\mathrm{[CII]}\Delta V \;\mathrm{and}\\
L^\prime_\mathrm{[CII]}&=&\frac{c^2}{2k_\mathrm{B}}\nu_\mathrm{obs}^{-2} D_L^2 S_\mathrm{[CII]}\Delta V(1+z)^{-3}
\end{eqnarray} 
for the total line flux \citep{Solomon1997} and listed in Tab.~\ref{tab:results}. 
A standard calibration uncertainty of 15\% for the ALMA Band-7 receiver\footnote{\url{https://almascience.nao.ac.jp/documents-and-tools/cycle6/alma-proposers-guide}} is included in the measurement uncertainties.

\section{Results and Analysis}\label{sec:results}

\begin{deluxetable*}{lcccccccccc}
\tablenum{2}
\tablecolumns{11}
\tabletypesize{\scriptsize}
\tablecaption{Results for the extended sources in SPT2349$-$56 that are discovered in [C\,{\sc ii}]158$\mu$m line emission.
\label{tab:results}}
\tablehead{
\colhead{ID} & 
\colhead{R.A. \& Declination} & 
\colhead{$A_\mathrm{spine}$} & 
\colhead{$D_\perp$} & 
\colhead{$S_\mathrm{[CII]}\Delta V$} & 
\colhead{$L_\mathrm{[CII]}$} & 
\colhead{$L'_\mathrm{[CII]}$} & 
\colhead{$\Delta v_\mathrm{[CII]}$} & 
\colhead{$\mathrm{FWHM}_\mathrm{[CII]}$} & 
\colhead{$M_\mathrm{gas}^{\mathrm{[CII]}}$} &
\colhead{$\Sigma_\mathrm{gas}^{\mathrm{[CII]}}$} \\
        & 
(J2000) & 
[kpc$^2$] & 
[kpc] & 
$\left[\frac{\mathrm{Jy\,km}}{\mathrm{s}}\right]$ & 
$\left[10^{8} L_{\sun}\right]$ & 
$\left[\frac{10^9\mathrm{K\,km\,pc}^{2}}{\mathrm{s}}\right]$ & 
$\left[\frac{\mathrm{km}}{\mathrm{s}}\right]$ & 
$\left[\frac{\mathrm{km}}{\mathrm{s}}\right]$ & 
[10$^8M_\sun$] &
$\left[\frac{M_{\sun}}{\mathrm{pc}^{2}}\right]$
}
\decimalcolnumbers
\startdata																		
S1(C16)  &  23:49:42.73--56:38:25.8  &  86.9  &  6.2  &  1.26$\pm$0.20  &  7.53$\pm$1.17  &  3.44$\pm$0.53  &  235$\pm$6  &  304$\pm$13  &  22.2$\pm$4.1  &  26$\pm$5  \\
S2(C23)  &  23:49:42.63--56:38:26.8  &  47.3  &  13.2  &  0.37$\pm$0.06  &  2.21$\pm$0.36  &  1.01$\pm$0.17  &  118$\pm$5  &  136$\pm$11  &  6.5$\pm$1.3  &  14$\pm$3  \\
S3(C22)  &  23:49:42.52--56:38:27.4  &  81.6  &  19.0  &  0.62$\pm$0.10  &  3.72$\pm$0.58  &  1.70$\pm$0.27  &  165$\pm$2  &  95$\pm$5  &  11$\pm$2.1  &  13$\pm$3  \\
S4       &  23:49:42.48--56:38:25.6  &  100.7  &  16.3  &  0.49$\pm$0.08  &  2.91$\pm$0.48  &  1.33$\pm$0.22  &  254$\pm$4  &  98$\pm$8  &  8.6$\pm$1.7  &  9$\pm$2  \\
S5       &  23:49:42.69--56:38:27.7  &  50.6  &  15.6  &  0.25$\pm$0.04  &  1.48$\pm$0.26  &  0.68$\pm$0.12  &  9$\pm$6  &  132$\pm$16  &  4.4$\pm$0.9  &  9$\pm$2  \\
S6       &  23:49:42.54--56:38:24.5  &  58.2  &  13.8  &  0.28$\pm$0.05  &  1.64$\pm$0.27  &  0.75$\pm$0.12  &  326$\pm$3  &  77$\pm$7  &  4.8$\pm$0.9  &  8$\pm$2  \\
S7       &  23:49:42.68--56:38:24.0  &  49.2  &  9.3  &  0.35$\pm$0.06  &  2.11$\pm$0.38  &  0.96$\pm$0.17  &  329$\pm$11  &  218$\pm$25  &  6.2$\pm$1.3  &  13$\pm$3  \\\tableline
CS1      &  23:49:42.89--56:38:24.2  &  36.3  &  5.4  &  0.67$\pm$0.15  &  3.99$\pm$0.89  &  1.82$\pm$0.40  &  -122$\pm$14  &  741$\pm$135  &  11.8$\pm$2.9  &  32$\pm$8 \\
CS2      &  23:49:42.93--56:38:24.5  &  26.3  &  7.7  &  0.41$\pm$0.07  &  2.44$\pm$0.42  &  1.12$\pm$0.19  &  -34$\pm$13  &  798$\pm$146  &  7.2$\pm$1.4  &  27$\pm$5  \\
CS3      &  23:49:43.04--56:38:25.1  &  41.1  &  13.6  &  0.36$\pm$0.11  &  2.16$\pm$0.68  &  0.99$\pm$0.31  &  72$\pm$17  &  264$\pm$40  &  6.4$\pm$2.1  &  16$\pm$5  \\
\tableline
S1-peak  &  23:49:42.73--56:38:25.8  &  11.1  &  6.2  &  0.37$\pm$0.06  &  2.23$\pm$0.34  &  1.02$\pm$0.53  &  267$\pm$4  &  203$\pm$8  &  6.6$\pm$1.2  &  59$\pm$11  \\
S2-peak  &  23:49:42.63--56:38:26.8  &  11.1  &  13.2  &  0.15$\pm$0.03  &  0.92$\pm$0.15  &  0.42$\pm$0.17  &  125$\pm$4  &  116$\pm$9  &  2.7$\pm$0.5  &  24$\pm$5  \\
S3-peak  &  23:49:42.52--56:38:27.4  &  11.1  &  19.0  &  0.24$\pm$0.04  &  1.45$\pm$0.22  &  0.66$\pm$0.27  &  166$\pm$2  &  73$\pm$3  &  4.3$\pm$0.8  &  38$\pm$7  \\
S4-peak  &  23:49:42.48--56:38:25.6  &  11.1  &  16.3  &  0.13$\pm$0.02  &  0.79$\pm$0.13  &  0.36$\pm$0.22  &  270$\pm$3  &  87$\pm$8  &  2.3$\pm$0.5  &  21$\pm$4  \\
S5-peak  &  23:49:42.69--56:38:27.7  &  11.1  &  15.6  &  0.14$\pm$0.02  &  0.81$\pm$0.13  &  0.37$\pm$0.12  &  4$\pm$4  &  107$\pm$9  &  2.4$\pm$0.5  &  21$\pm$4  \\
S6-peak  &  23:49:42.54--56:38:24.5  &  11.1  &  13.8  &  0.11$\pm$0.02  &  0.63$\pm$0.11  &  0.29$\pm$0.12  &  333$\pm$3  &  76$\pm$8  &  1.9$\pm$0.4  &  17$\pm$3  \\
S7-peak  &  23:49:42.68--56:38:24.0  &  11.1  &  9.3  &  0.18$\pm$0.03  &  1.04$\pm$0.18  &  0.48$\pm$0.17  &  286$\pm$11  &  250$\pm$24  &  3.1$\pm$0.6  &  27$\pm$6  \\\tableline
CS1-peak  &  23:49:42.89--56:38:24.2  &  11.1  &  5.4  &  0.49$\pm$0.16  &  2.93$\pm$0.94  &  1.34$\pm$0.40  &  -99$\pm$22  &  575$\pm$74  &  8.6$\pm$2.9  &  77$\pm$26  \\
CS2-peak  &  23:49:42.93--56:38:24.5  &  11.1  &  7.7  &  0.22$\pm$0.05  &  1.33$\pm$0.28  &  0.61$\pm$0.19  &  -42$\pm$8  &  657$\pm$124  &  3.9$\pm$0.9  &  35$\pm$8  \\
CS3-peak  &  23:49:43.04--56:38:25.1  &  11.1  &  13.6  &  0.15$\pm$0.03  &  0.90$\pm$0.17  &  0.41$\pm$0.31  &  131$\pm$12  &  227$\pm$28  &  2.7$\pm$0.6  &  24$\pm$5  \\
\tableline
$\int$\,Ext.  &  ---   &  578  &  ---  &  5.06$\pm$0.33  &  30.2$\pm$1.9  &  13.8$\pm$0.9  &  135$\pm$3  &  286$\pm$20.6  &  89.1$\pm$6.6  &  15.4$\pm$1.1  \\
\tableline
\enddata
\tablecomments{$^{(1)}$Source IDs from \citet{Hill2020} that are co-spatial with new candidates are indicated in brackets. $^{(2)}$Co-ordinates refer to the $M_0$ peak pixel position within each segment. 
$^{(4)}$For the segments 'S1' through 'S7' the projected physical distance to SMG 'C' and for the counter-streamer segments 'CS1' through 'CS3', the projected physical distance to SMG 'B' is tabulated. 
$^{(8)}$Line-of-sight velocity $\Delta v_\mathrm{[CII]}$ relative to systemic redshift $z=4.303$ is measured from the spectral line fit (see Sec. \ref{sec:reduction}). For segments 'CS1' and 'CS2' where a double-Gaussian component is required, $\Delta v_\mathrm{[CII]}$ is given for the center of the narrow component only. 
$^{(9)}$For IDs fitted by a double-Gaussian model, the line width refers to the broader width. 
$^{(11)}$Gas surface density $\Sigma_\mathrm{gas}^\mathrm{[CII]}$ is calculated from the gas mass measured within the segment (peak) divided by $\Omega_\mathrm{spine}$ ($\Omega_\mathrm{bm}$).
}
\end{deluxetable*}

In a previous study on cold gas in SPT2349$-$56, extended [C\,{\sc ii}] emission around the central ULIRG triplet was discovered  \citep{Hill2020}. Because we are interested in the physical mechanisms that take place in this region, we show the moment-0 (first row), moment-1 (second row) and moment-2 (third row) of the southern LABOCA source of SPT2349$-$56 (left column) and a 180$\times$150\,kpc$^2$ zoom-in (right column) in Fig.\ref{fig:map}. The region is populated by 14 DSFGs \citep{Miller2018}, labels from `A' to `N' according to their $S_{850}$ brightness. Although \citep{Hill2020} ranked new-found line emitters by their [C\,{\sc ii}] luminosity and classified `C16', `C23', and `C22' as galaxy candidates (`S1', `S2', `S3', respectively), our deeper spectral imaging shows that the three sources are coherently connected by diffuse [C\,{\sc ii}] emission. Thus, we now classify these features not part of any galaxy, but extended emission ``clumps'' with relatively bright peak intensities located in the circumgalactic medium (see also Fig.~\ref{fig:momentszoom}). Two distinct extended structures are found: a loop to the south-west of the central SMGs `B'-`C'-`G' called the ``western streamer'' and an elongated extension to the south-east of SMG `B' named the ``counter-streamer''. The western streamer seems to fragment into seven distinct, yet coherently connected [C\,{\sc ii}]-emitting regions (`S1'--`S7'). While the counter-streamer is segmented into three regions (`CS1'--`CS3'). In addition to the [C\,{\sc ii}] streamers, 11 new galaxy candidates were detected in the deep moment maps (see Sec.~\ref{appendix:newgalaxies}).

\subsection{Bright [CII] streamers}\label{sec:lines}

The western streamer segments add up to 0.26\,Jy\,km\,s$^{-1}$ or $ L_\mathrm{[CII]}=21.6\pm1.5\times10^8$\,$L_\sun$ and have $S/N>10$ each. The combined peak line flux density is $\sim$12\,mJy. ULIRG `G' connects to the south with segment `S1' and SMG `B' overlaps with segment `S7'. `S3', projected at a distance of $D_\perp\sim33$\,kpc from `C' marks the distant tip of the arc. Our more sensitive measurements show a 3$\times$ enhancement in the specific intensity of source `S3' with 3.2$\pm$0.5\,mJy\,arcsec$^{-2}$ over the value reported in \citet{Hill2020}. 

We approximate the sizes of flux detection as circles with areas $A_\mathrm{spine}=\pi R_\mathrm{spine}^2$, referred to as ``spines'', with respective circularized spine radii of 2.9--5.7\,kpc (not de-convolved with the beam). Typical FWHM sizes of these [C\,{\sc ii}] clumps are roughly $\sim$5\,kpc, thus only marginally larger than the synthesized beam and only slightly larger than the half-light radii of DSFGs of comparable $L_\mathrm{[CII]}$ in the system \citep{Hill2020}. The thin and clumpy morphology of the western streamer is evokative of ``beads on a string'' \citep{Elmegreen2007}. 

The velocity dispersion of the sources range between $\sigma_v=129\pm6$\,km\,s$^{-1}$($86\pm4$\,km\,s$^{-1}$) for `S1'(`S1-peak') to $\sigma_v=33\pm3$\,km\,s$^{-1}$($32\pm3$\,km\,s$^{-1}$) for `S6'(`S6-peak'). `S3-peak' sticks out with $\sigma_v=31\pm1$\,km\,s$^{-1}$, the lowest velocity dispersion, despite showing the second highest peak-intensity of the western segments. This structure is therefore characterized by high line intensities with relatively low velocity dispersions and single-component Gaussian fits well reproduce the [C\,{\sc ii}] velocity profiles of these clumps (see Fig.~\ref{fig:spectra}).

Connected to SMG `B' is the counter-streamer, detected at a similar velocity offset to the western segments, but with a stronger negative velocity component, extending below $\Delta v\lesssim-100$\,km\,s$^{-1}$. The counter-streamer measures a total of $S_\mathrm{[CII]}\Delta V=1.44\pm0.20$\,Jy\,km\,s$^{-1}$ or $L_\mathrm{[CII]}=8.6\pm1.2\times10^8$\,$L_\sun$. The sky separation between `B' and `CS3', at the tip of the counter-streamer, is $\sim$1.97$\arcsec$ or $D_\perp=13.6$\,kpc. While `CS1' to `CS2' seem to be smoothly connected, `CS3' shows more complex morphology. `CS1' and `CS2' require two-component Gaussian models to accurately recover the [C\,{\sc ii}] velocity profiles with a broad and more narrow component. While the narrow component's FWHM$_\mathrm{[CII]}$ are similar to values found in the western streamer at $\sim110$--$160$\,km\,s$^{-1}$, the broad components at $\mathrm{FWHM}_\mathrm{[CII]}=750$--$800$\,km\,s$^{-1}$ are likely caused by insufficient subtraction of the extended [C\,{\sc ii}] morphology close to SMGs `B' and `C', overlapping in velocity range. Given our limited spatial resolution, a contribution from nuclear outflowing material associated with the bright radio source `C' \citep{Chapman2024}, can currently not be fully excluded.

Combining the western streamer and counter-streamer sources, we find a total of $S_\mathrm{[CII]}\Delta V = 5.6\pm0.33$\,Jy\,km\,s$^{-1}$ or $3.02\pm0.19\times10^9$\,$L_\sun$ of extended [C\,{\sc ii}] luminosity. The total line luminosity of the extended emission is similar to that of HyLIRG 'C' \citep{Miller2018,Hill2020}. Importantly, the combined [C\,{\sc ii}] peak flux density of the western streamer is as high as the peak flux density of HyLIRG 'B', with the counter-streamer's integrated peak flux density rivaling that of ULIRG 'G' (see Fig.~\ref{fig:spectra}).

\subsection{Morpho-kinematic links to the ULIRG triplet}\label{sec:gradients}

It appears as if the extended [C\,{\sc ii}] emission mirrors the main spectral features of the massive\footnote{Stellar masses of `B' and `G' are so-far undetermined. `C' is bright in rest-frame UV and yields $M_\ast\approx10.9\pm 3.4\times10^{10}\,M_\sun$ \citep{Hill2022}. The gas mass ratio of $\mu_\mathrm{gas}=0.31$ predicts a baryonic mass of $(1+\mu_\mathrm{gas})M_\ast\approx14.3\times10^{10}\,M_\sun$, comparable to that of ultra-massive galaxies \citep{Forrest2020}. Importantly, 'C' hosts a radio-loud and X-ray-bright AGN \citep{Chapman2024,Vito2024}.} protocluster galaxies, despite the relatively large projected distances ($>$10\,kpc) between the different sources.
Fig.~\ref{fig:spectra} shows a gallery of spatially integrated spectra of SMGs `B'-`C'-`G' and DSFG `C12' (top) and [C\,{\sc ii}] line profiles extracted from the individual streamer segments (left column) and peak intensity positions (right column). The integrated spectrum of the western streamer fills the velocities between the blue- and red-wing of galaxies `C' and `B', while covering a similar velocity range as the broad, double-horn profile of `G' (see Fig.~\ref{fig:spectra}). 
It appears as if the extended [C\,{\sc ii}] emission mirrors the main spectral features of the massive protocluster galaxies, despite the relatively large projected distances ($>$10\,kpc) between the different sources.

Galaxies `B'-`C'-`G' show velocity gradients consistent with disk-like rotation \citep{Venkateshwaran2024} and projected major axes roughly aligning east-west -- albeit `C' would be considered counter-rotating with respect to `B' and `G' \citep{Chapman2024}. Assuming that the three-dimensional distances are comparable to the projected distances at $D_\perp<15$\,kpc, the triplet is likely undergoing a multiple-component major\footnote{We find $L^\mathrm{G}_\mathrm{[CII]}/L^\mathrm{B}_\mathrm{[CII]}\approx1$:$4.3$ and $L^\mathrm{G}_\mathrm{[CII]}/L^\mathrm{C}_\mathrm{[CII]}\approx1$:$2.8$.} merger. 
Also, we see an extended $\sim$5\,kpc gaseous structure\footnote{The centroid position of the `C'--`G' bridge, estimated from the dust continuum map, is 23$^\mathrm{h}$49$^\mathrm{m}$42.79$^\mathrm{s}-$56$\degr$38$\arcmin$24.9$\arcsec$ (J2000).} connecting `G' and `C' consistent with a tidal bridge \citep{Toomre1972}.

Source `C12' is roughly co-spatial with `CS1', $D_\perp=5$\,kpc to the east of 'B' and $\sim$630\,km\,s$^{-1}$ offset from the narrow component of `CS1'. No apparent bridge or tidal features is detected between `C12'--`CS1'. It is currently unclear whether source `C12' is tidal debris or a galaxy, forming a quartet with SMGs `B'-`C'-`G'.

Due to projection effects, galaxies `L' and `LBG3' are found in proximity to `S4' and `S5', respectively. While a past, high-velocity interactions between the western streamer and 'L' or `LBG3' are conceivable, the velocity offset of $\sim$660\,km\,s$^{-1}$ and $\sim$1000\,km\,s$^{-1}$ are substantial.

\begin{figure*}
\epsscale{1.0}
\center
\includegraphics[width=0.99\textwidth]{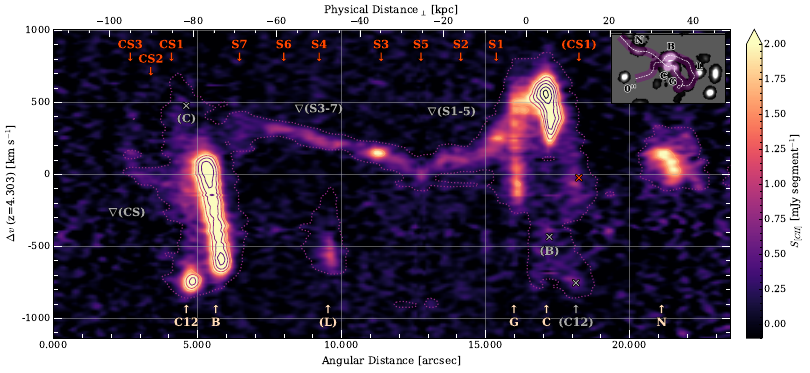}
\caption{Position--velocity diagram extracted along the curved spines of the [C\,{\sc ii}]-streamers. Inset shows the $M_0$ map with the ULIRG triplet along the looped extraction path as the white dashed curve. The extraction height is 0.75$\arcsec \approx 5$ kpc (purple curve in inset), corresponding to a segment area of $0.75\arcsec\times0.1\arcsec$. In red, the positions of the streamer peak positions are marked (top); in peach (bottom), SMGs positions that fall into the segment height are marked. The origin of the projected distance $D_\perp$ is set at the ionized carbon bridge between `C' and `G' with $+500$ km s$^{-1}$. Dotted purple contour gives the outer extended of the 2-$\sigma$ clipped cube (see section \ref{sec:reduction}). [C\,{\sc ii}] sources in brackets are only partially covered by the height of the extraction segment. The ``$\times$''-markers denote flux that is (partially-) doubled due to the looped extraction path. Note the strikingly coherent streaming structure `$\nabla$(S3-7)', including `S5', that extends linearly between `(C)' and 0\,km\,s$^{-1}$.
\label{fig:pv}}
\end{figure*}

Overall, the streamers show complex morphology on the sky. A detailed analysis of the velocity-structure, utilizing position-velocity (PV) diagrams, shown in Fig.~\ref{fig:pv}, allows to further probe the morpho-kinematics of the extended [C\,{\sc ii}] emission. The extraction is performed on a curved extraction path to trace the arc-like structures\footnote{We utilize the Python tool \texttt{pvextractor} \url{https://github.com/radio-astro-tools/pvextractor} in version 0.2 \citep{Ginsburg2016} with a custom spline-interpolation for the extraction path.}.  The combined length along the path is $l_\perp=60$\,kpc, spanning between `G' and `B', and split by two straight velocity gradients, $\nabla$(S1-5) and $\nabla$(S3-7).

It is conceivable that the western streamer is not a single coherent structure in 3-D, but consists of two projected streams\footnote{Alternatively, the arcs can be further decomposed into several coherent structures, as shown in Fig.~\ref{fig:additional-pv}.}, with `S5' marking a kinematic watershed. To one side, between `S1' and `S2', the velocity gradient seems to be steep with $\nabla\mathrm{(S1-2)}\approx16.7$\,km\,s$^{-1}$\,kpc$^{-1}$ and by extrapolation, appears to stretch towards the red-part of `C', while to the other side, a remarkably consistent gradient over $l_\perp=45$\,kpc with $\nabla\mathrm{(S3-7)}=8.0$$\pm$$1.6$\,km\,s$^{-1}$\,kpc$^{-1}$ is found.

At the position of `S5', close to $\Delta v=0$\,km\,s$^{-1}$, we possibly see the western streamer looping back onto itself \citep[cf.][]{Bournaud2011}. For this reason, Fig.~\ref{fig:pv} includes a loop towards `S5'. Also, a slight enhancement in velocity dispersion between `S2' and `S5' might signify the aforementioned line-of-sight projection of different velocity components \citep{Bournaud2004}. 

The morpho-kinematics of the counter-streamer appears to mirror the velocity intersection of the outer disk of galaxy `B' and the blue part of `G'. Measured on the narrow Gaussian components, the counter-streamer shows a general steep gradient $\nabla\text{(CS1-CS3)} = 22.0$$\pm$$1.7$ km s$^{-1}$ kpc$^{-1}$ with `CS3' being broad with $\mathrm{FWHM}_\mathrm{[CII]}=264\pm40$\,km\,s$^{-1}$, corroborating a line-of-sight loop-like feature, seen in $M_1$ and $M_2$.

\subsection{Dust continuum detection}\label{sec:continuum_model}

\begin{figure*}[ht!]
\centering
\epsscale{1.0}
\includegraphics[width=0.999\textwidth]{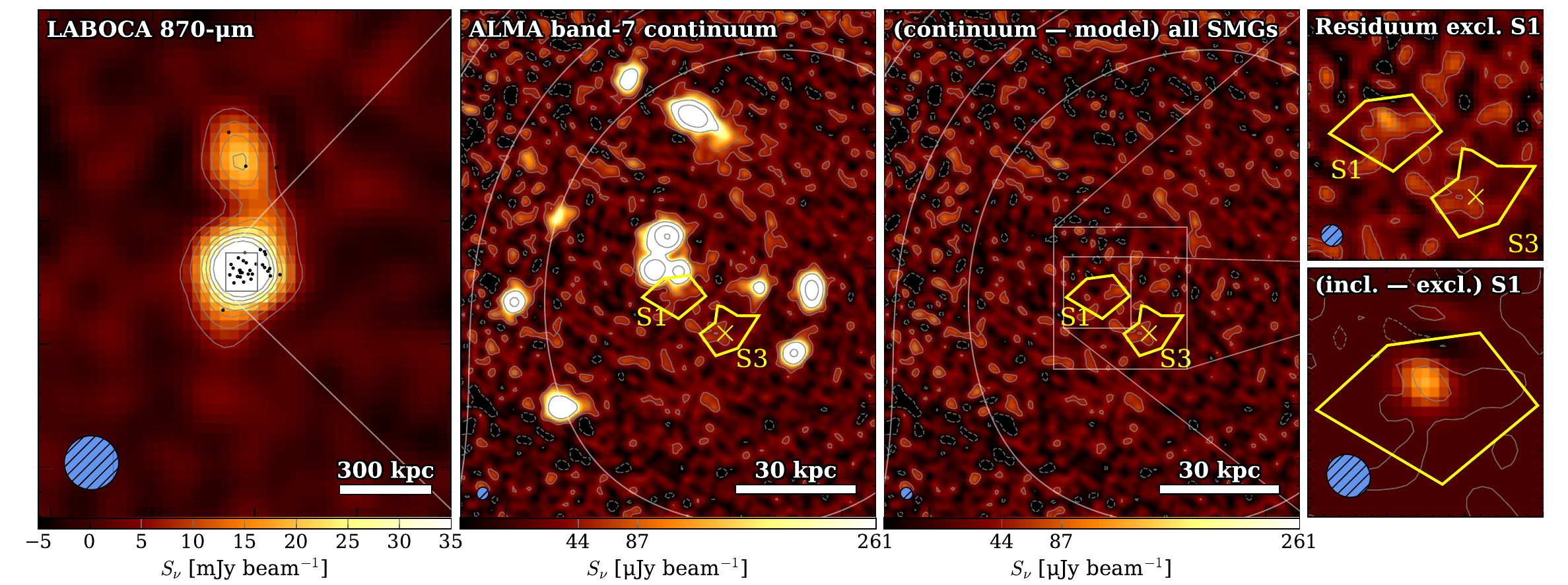}
\caption{Dust continuum maps of SPT2349$-$56 in the sub-millimeter and at different spatial scales. The protocluster core fragments into a north- and south-component (left panel) on the APEX/LABOCA map \citep{Hill2020}. Black dots denote individual [C\,{\sc ii}]-source positions detected with ALMA. The zoom-in on the south component shows the ultra-deep combined ALMA Band-7 continuum image at 346.41\,GHz and [-3,3,6,18,72,120,360]$\times\sigma$ contours for $\sigma=14.9$\,$\mu$Jy\,beam$^{-1}$, corrected for the primary beam illumination at source `S1'; see also Fig.~\ref{fig:map}). The flux density cuts are consistent among all Band-7 images. Note that there is no compact continuum counter-part detected for the bright [C\,{\sc ii}] flux peak within segment `S3' (yellow ``$\times$''). 
\label{fig:continuum}}
\end{figure*}

The source and residual image, after subtracting all ALMA Band-7 continuum models (see Sec.~\ref{sec:continuum}) are shown in the center and right panel in Fig.~\ref{fig:continuum}. Overall, the [C\,{\sc ii}] streamers are undetected in 850$\mu$m dust continuum, with the exception of the [C\,{\sc ii}]-brightest segment `S1' to which we detect an unresolved continuum counterpart at a signal-to-noise level of $S/N=10.5$ (see zoom-in panels to the right). The flux density of continuum source `S1' is $S_{850}=0.15\pm0.03$\,mJy; applying a 15\% calibration uncertainty for Band-7 \citep{Bonato2018,Francis2020}. 

The inclusion of the [C\,{\sc ii}] bridge component with $S/N=17$ into the six-component continuum model is necessary to minimize the flux density residuum and keep the overall systematic uncertainties low for weaker component `S1', just $\sim$0.9$\arcsec$ to the south of the bridge. For the `C'--`G' bridge, a flux density of $S_{850}=0.25\pm0.04$\,mJy is obtained. 

Because the continuum of streamer segment `S1' is mostly blended with the extended continuum of galaxy `G', the systematic uncertainties of the best-fitting continuum model for `S1' are intrinsically high. We acknowledge that the continuum image still shows patchy excess residuals from the \texttt{CLEAN} process. Nevertheless, the positional offset of the continuum counterpart to the peak [C\,{\sc ii}] intensity within segment `S1' is $\Delta\theta=0.17\arcsec$, well below the synthesized beam size.

A tentative $\lesssim$5$\sigma$ extended signal is found to the east of segment `S3', but not co-spatial with the peak of [C\,{\sc ii}] emission. Based on the assumption that [C\,{\sc ii}] emission is typically more extended than the dust continuum \citep{Venemans2020}, and to account for non-detections, we scale the background RMS noise to obtain 3$\sigma$ upper limits. 
Continuum non-detections are then calculated as $S_\mathrm{850}<3\sigma_{850,\mathrm{spine}}=3\times\sqrt{{\Omega_\mathrm{spine}}/{\Omega_\mathrm{bm}}}({\sigma_{850}^{S/N}}/{f_\mathrm{pb})}$ with $\Omega_\mathrm{spine}$ as the solid angle of the line detection and $\Omega_\mathrm{bm}=0.233$\,arcsec$^2$ for the synthesized beam area under the small angles approximation. For `S3', we obtain $S_\mathrm{850}<0.11$\,mJy, or $\sim$0.75 times the continuum flux density of `S1'. This tentative continuum upper limit is consistent with $L_\mathrm{[CII]}^\mathrm{S1}/L_\mathrm{[CII]}^\mathrm{S3}\approx0.5$ supporting shared cold gas properties among `S1' and `S3'.

The counter-streamer source `CS1', although as bright in [C\,{\sc ii}] as `S3', is not detected in continuum. However, this segment is heavily blended with the continuum counterpart of `C12'. All dust continuum measurements and upper limits are listed in Tab.~\ref{tab:results_continuum}.

\subsection{Cold gas mass contained in the streamers}\label{sec:mass}\label{sec:gasmass}

The physical properties of vastly extended gas reservoirs around SMGs, traced by [C\,{\sc ii}], are largely unexplored at $z>2$ \citep{Hollenbach1991,Madden2020,Dessauges-Zavadsky2020,Herrera-Camus2021,Vizgan2022,Zhou2025}. \citet{Zanella2018} reported that the [C\,{\sc ii}] line luminosity in star-forming main-sequence galaxies at $z\sim2$ scales linearly with the carbon monoxide (CO) derived molecular gas mass. The relationship is tight with $\pm$0.3 dex scatter over three orders of magnitude in $L_\mathrm{[CII]}$. 
Main sequence galaxies in the distant Universe follow a normalization coefficient of $\alpha_\mathrm{[CII]}^\mathrm{Z18}=M_\mathrm{mol}/L_\mathrm{[CII]}=31^{+31}_{-16}$\,$M_\sun$\,$L_\sun^{-1}$ apparently independent of star-formation rate, redshift, or metallicity \citep{Zanella2018}. The strongly lensed and highly star-forming SPT-SMG sample from \citet{Gullberg2015} shows a slightly lower average conversion factor with $\alpha_\mathrm{[CII]}=22$\,$M_\sun\,L_\sun^{-1}$ although consistent within 0.3\,dex uncertainty \citep[see][]{Gururajan2023}.

For a minimum mass limit of the [C\,{\sc ii}] streamers, we use the formula from \citet{Venemans2017} that assumes a brightness temperature of C$^+$ ions in the optically thin limit and relates it to the average surface intensity by 
\begin{eqnarray}
&M_\mathrm{C^{+}}&=2.92\times10^{-4}M_\sun\frac{Q(T_\mathrm{ex})}{4}e^{91.2/T_\mathrm{ex}}L_\mathrm{[CII]}^{\prime}\;\mathrm{and}\\
&M_\mathrm{gas}^{\mathrm{C}^+}&\geq\frac{(1+f_\mathrm{He})m_\mathrm{H}}{m_{C}}([\mathrm{C}]/[\mathrm{H}])^{-1}f_\mathrm{[CII],ion}^{-1}\times M_\mathrm{C^{+}}\label{eq:venemans}
\end{eqnarray} 
with the partition function $Q(T_\mathrm{ex}=100\,\mathrm{K})$. The minimum cold gas mass is then derived by assuming that all carbon atoms are singly ionized, a cosmic helium fraction $f_\mathrm{He}=0.36$, and solar carbon abundance $[\mathrm{C}]/[\mathrm{H}]=2.69\times10^{-4}$ \citep{Asplund2009}, yielding $M_\mathrm{gas}^{\mathrm{C}^+}=9.49\times10^8\,M_\sun$ for 'S1', or an equivalent $\alpha_\mathrm{[CII]}\approx1.3\,M_\sun\,L_\sun^{-1}$. 
This means that the total theoretical minimum masses (Eq.~\ref{eq:venemans}) and empirically calibrated cold gas masses calculated with $\alpha_\mathrm{[CII]}^\mathrm{Z18}=31\,M_\sun\,L_\sun^{-1}$ from \citet{Zanella2018} disagree by a factor of $\sim$24 for the same [C\,{\sc ii}] luminosity. 

\begin{figure}[ht!]
\flushright
\epsscale{1.0}
\includegraphics[width=0.47\textwidth]{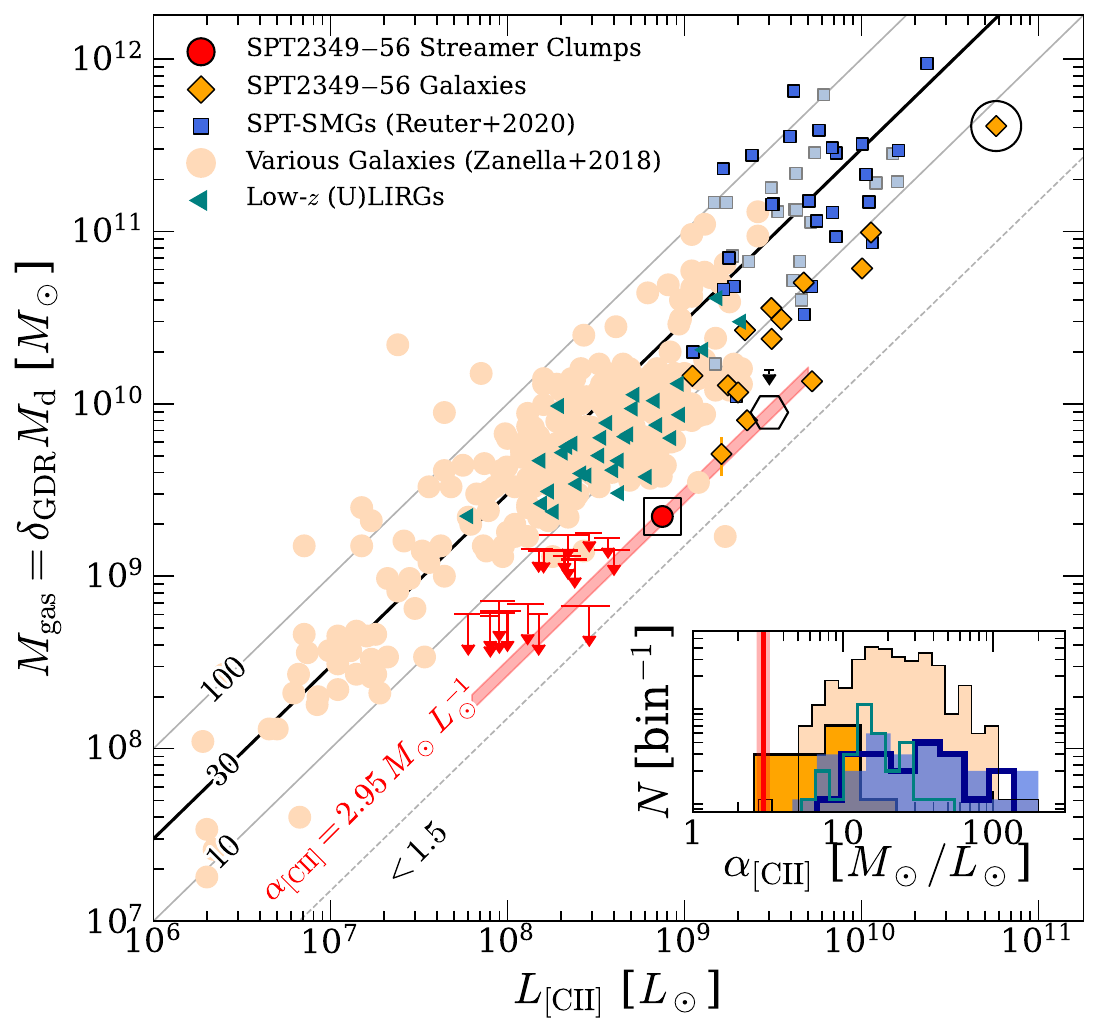}
\caption{Inferred $\alpha_\mathrm{[CII]}$ (diagonal curves; in units of $M_\sun\,L_\sun^{-1}$) for various of low- and high-$z$ galaxies (peach dots) from \citet{Zanella2018}. A constant $\delta_\mathrm{GDR}=100$ is employed for lensed SPT-SMGs (\citealt{Reuter2020}; using either individual corrections for source magnification or $\langle \mu \rangle=5.5$ otherwise; light blue squares), \citet{Hill2020} SMGs in SPT2349$-$56, and the streamer clumps. The summed $\delta_\mathrm{GDR}M_\mathrm{d}$ and $L_\mathrm{[CII]}$ of all SMGs in SPT2349$-$56 (assuming $T_\mathrm{d}=39.6$\,K) is indicated by the encircled orange diamond. `S1' is marked by a black box and the open hexagon (black arrow) marks the combined $L_\mathrm{[CII]}$ and inferred gas mass (3$\sigma$ upper limit for $\delta_\mathrm{GDR}M_\mathrm{d}$) of the streamers. The inlet shows low-$z$ (U)LIRGs from \citet{Herrero-Illana2019} (cyan histogram) and the \citet{Hill2020} galaxies (orange histogram) having systematically lower $\alpha_\mathrm{[CII]}$. The red line marks the low $\alpha_\mathrm{[CII]}$ and $\pm$1$\sigma$ standard deviation of `S1'.
\label{fig:alphacii}}
\end{figure}

Alternatively, a \emph{concordance gas mass} can be established based on a dust mass ($M_\mathrm{d}$) estimate. The cold gas mass then scales linearly by $M_\mathrm{gas}=\delta_\mathrm{GDR} M_\mathrm{d}$ assuming a constant molecular gas-to-dust mass ratio of $\delta_\mathrm{GDR}=100$ for simplicity. In Fig.~\ref{fig:alphacii}, dust-based (molecular) gas estimates are compared to $L_\mathrm{[CII]}$. Since only the continuum component of `S1' is detected, we extrapolate dust masses for all segments based on this measurement. 

In order to obtain dust masses, the method from \citet{Weiss2007} is adopted to directly fit the flux density measurement to a modified blackbody model \begin{equation}\label{equ:sobs} S_\mathrm{obs}(\nu_\mathrm{obs};M_\mathrm{d})=\frac{(1+z)\pi R_\mathrm{gal}^2}{D_L^2} B'_\nu(T_\mathrm{d}) \left[1-\mathrm{e}^{-\frac{M_d\kappa(\nu_0)}{\pi R_\mathrm{gal}^2}} \right]\end{equation} with the dust mass as the only free parameter\footnote{The Levenberg-Marquardt least-squares algorithm is employed via the Python function \texttt{curve\_fit} from the package \texttt{scipy.optimize} (\url{https://docs.scipy.org/doc/scipy/reference/generated/scipy.optimize.least squares.html}).}. The blackbody $B_{\nu}'(T_d)=[B_{\nu}(T'_\mathrm{d})-B_{\nu}(T_\mathrm{CMB}(z))]$ is corrected for the effect of the CMB heating the dust to $T'_\mathrm{d}$ as a function of redshift and decreasing the contrast to the warmer background \citep[see e.g.][]{Jones2020}. Further, the dust absorption coefficient of $\kappa(\nu_0)=\kappa_0(\nu_0/250\,\mathrm{GHz})^\beta$ with $\kappa_0=0.045$ m$^2$ kg$^{-1}$ \citep{Hildebrand1983,Kruegel1994,Greve2012}, dust emissivity index $\beta=2$, and an equivalent, optically thick radius $R_\mathrm{gal}=\sqrt{\kappa(\nu_0) M_\mathrm{d}}$ is adopted. To account for optical depth, we use $\lambda_0=c/\nu_0=100$\,$\mu$m as the wavelength where the dust SED becomes optically thick. $\kappa(\nu_0)$ is then evaluated at the rest frequency $\nu_0=(1+z)\nu_\mathrm{obs}$ of the observations at redshift $z=4.303$.  
For `S1', a dust temperature $T_d=30$ K is assumed. \citet{Hill2020} found a weighted average dust temperature of $39.6$\,K among the SPT2349$-$56 SMGs that would lower the dust masses by $\sim$44\%. We assume a lower dust temperature since the dust in the extended structure might not be heated as efficiently by young stars. To account for the unconstrained parameter space, Tab.~\ref{tab:models} summarizes results for sets of dust properties within 16$^\mathrm{th}$--84$^\mathrm{th}$ percentiles as inferred from the single photometric point of `S1'. The specific models were chosen to demonstrate the variation in parameter uncertainties of $\{M_\mathrm{d}, T_\mathrm{d}, \lambda_0\}$ when these parameters are treated as constants for the fit. 

By additionally assuming an equivalent brightness temperature to a typical excitation temperature of [C\,{\sc ii}] emitting gas in the ISM of $T_\mathrm{ex}=100$\,K \citep[see e.g.][]{Salas2021} within the area of `S1' with $A_\mathrm{spine}=86.9$\,kpc$^2$, an effective line emitting area $A_\mathrm{[CII]}=A_\mathrm{spine}T_\mathrm{b,[CII]}/T_\mathrm{ex}\approx0.113$\,kpc$^2$ is predicted\footnote{The [C\,{\sc ii}] brightness temperature of segment `S1' is calculated via the Rayleigh-Jeans approximation $T_\mathrm{b,[CII]}=(1+z)c^2S_\mathrm{[CII]}/(2 k_B\nu_{obs}^2\Omega_\mathrm{spine})$ with $S_\mathrm{[CII]}=S_\mathrm{[CII]}\Delta V/\mathrm{FWHM}_\mathrm{[CII]}$. For `S1', we find $T_\mathrm{b,[CII]}=0.130$\,K and a corresponding beam filling factor of $\Phi_A\approx T_\mathrm{b,[CII]}/T_\mathrm{ex}=0.0013$.}. \texttt{Model-A} in Tab.~\ref{tab:models} couples the dust size $\pi R_\mathrm{gal}^2$ to the inferred physical size of the [C\,{\sc ii}] emitting area $A_\mathrm{[CII]}$ of 'S1' at a constant concordance dust mass. Coupling the dust to [C\,{\sc ii}] emitting size, we conclude that even a low $T_\mathrm{d}=30$\,K allows the dust to be warm enough to emit the observed flux density within $A_\mathrm{[CII]}\approx0.113$\,kpc$^2$. 

Assuming identical dust properties throughout the [C\,{\sc ii}] streamer segments, a concordance mass-to-light ratio of $\alpha_\mathrm{[CII]}=2.95\pm0.30$ for the extended gas is found and marked as the red curve in Fig.~\ref{fig:alphacii}. 
According to our conversion factor, the combined gas mass of the streamers amounts to $89.1\pm6.6\times10^8$\,$M_\sun$ (see black dot in Fig.~\ref{fig:alphacii}). Within full model uncertainties, we find a combined gas mass $M_\mathrm{gas}=1.16^{+1.43}_{-0.60}\times 10^{10}$\,$M_\sun$ and combined star-formation rate $\mathrm{SFR}_\mathrm{FIR}=7.7^{+12}_{-4.4}$\,$M_\sun$\,yr$^{-1}$ by scaling the values of \texttt{Model-3} in Tab.~\ref{tab:models} to the relative contribution of `S1' ($\sim$24.9\%) to the combined $L_\mathrm{[CII]}$.

The theoretically estimated minimum gas masses for all clumps (see Eq.~\ref{eq:venemans}) are systematically lower than the $T_\mathrm{d}=30$\,K dust-inferred masses by a factor of $\sim$2--3. Slightly higher dust temperatures, as inferred for the SMGs in the protocluster core \citep{Hill2020}, can not be fully rejected. However, for $T_\mathrm{d}\gtrsim40$\,K, the $M/L$-ratio for ‘S1’ would fall below the allowed range at solar metalicity ($\alpha_\mathrm{[CII]}<1.5\,M_\sun L_\sun^{-1}$; see \citealt{Herrera-Camus2021}). On the other hand, below $T_\mathrm{d}=24.5$\,K, while allowing larger $\alpha_\mathrm{[CII]}$ values, the average dust temperature would be less than 10\,K above the $z=4.3$ CMB temperature, atypical for star-forming gas.

Importantly, the $\alpha_\mathrm{[CII]}^\mathrm{Z18}$-masses are an order of magnitude higher than predicted by the concordance model. The molecular component of `S1' with $\alpha_\mathrm{[CII]}=31\,M_\sun\,L_\sun^{-1}$ would be as massive as the bright galaxy `C' and clump `S3' would be more massive than SMG `G'. Consequently, the gas mass of the extended emission would then exceed the combined gas mass of the central ULIRG triplet ($\gtrsim10^{11}$\,$M_\sun$) -- which appears implausible.

\begin{table}[h!]
\setlength{\tabcolsep}{2pt}
\scriptsize
\tablenum{3}
\center
\caption{Comparison of different dust continuum models, reproducing the measured flux density of `S1' with Eq.~\ref{equ:sobs}. Posterior medians, 16$^\mathrm{th}$, and 84$^\mathrm{th}$ percentiles are evaluated following a Markov Chain Monte Carlo simulation (50000 samples) with model-dependent priors: $M_\mathrm{d} \in\mathcal{U}(10^6,10^8)\times M_\sun$, $T_\mathrm{d}\in\mathcal{U}(T_\mathrm{CMB,0}(1+z)+10,70)$\,K, $\lambda_0 \in \mathcal{U}(3,300)$\,$\mu$m, $\beta=2$. ``$\diamond$'' indicates fixed priors for the concordance model (\texttt{C.~M.}) and values in brackets for \texttt{Model-A} are consistent with $T_\mathrm{ex}=100$\,K (see main text).
\label{tab:models} 
}
\begin{tabular}{lcccccc}
\tableline
\tableline
& $M_\mathrm{d}$ & $T_\mathrm{d}$ & $\lambda_0$ & $\pi R_\mathrm{gal}^2$ & $L_\mathrm{FIR}$ & $\mathrm{SFR}_\mathrm{FIR}$ \\
&  [$10^6\,M_\sun$] & [K]          & [$\mu$m]    & [kpc$^2$]        & [$10^{8}\,L_\sun$] & [$M_\sun$yr$^{-1}$] \\
\tableline
\texttt{C.~M.} & 22.2$\pm$3.9 & 30 & 100 & 0.30$\pm$0.05 & 269$\pm$47 &  2.6$\pm$0.4 \\
\tableline
\texttt{Model-A} & $\diamond$ & $32.6^{+2}_{-2}$ & ($163$) & ($0.113$) & $209^{+86}_{-107}$ & $2.0^{+0.8}_{-1.0}$ \\
\texttt{Model-0} & $\diamond$ & $35.6^{+3}_{-5}$ & $158^{+92}_{-98}$ & $0.12^{+0.54}_{-0.07}$ & $251^{+129}_{-93}$ & $2.4^{+1.2}_{-0.9}$ \\
\texttt{Model-1} & $\diamond$ & $29.9^{+2}_{-2}$ & $\diamond$ & $\diamond$ & $263^{+109}_{-93}$ & $2.5^{+1.0}_{-0.9}$  \\
\texttt{Model-2} & $20.0^{+17.2}_{-14.5}$ & $30.8^{+20}_{-5}$ & $\diamond$ & $0.27^{+0.23}_{-0.20}$ & $302^{+121}_{-147}$ &  $2.8^{+12}_{-1.4}$ \\
\texttt{Model-3} & $28.8^{+35.7}_{-15.0}$ & $32.3^{+11}_{-6}$ & $226^{+54}_{-100}$ & $0.11^{+0.14}_{-0.07}$ & $200^{+325}_{-116}$ & $1.9^{+3.1}_{-1.1}$ \\
\tableline
\end{tabular}
\end{table}

Lastly, the 3$\sigma$ upper limit for the molecular mass inferred from CO($J\,{=}\,4\,{-}\,3$) \citep{Hill2020,Rotermund2021} agrees well with the concordance mass. From the upper limit $L^\prime_\mathrm{CO(4-3)}<6.78\times10^8$\,K\,km\,s$^{-1}$\,pc$^2$, the molecular mass is calculated with $M_\mathrm{mol}<\alpha_\mathrm{CO}(1+f_\mathrm{He})L^\prime_\mathrm{CO(4-3)}/r_{4,1}$ \citep{Carilli2013} assuming $r_{4,1}=0.6$.  
For all cases, $\alpha_\mathrm{[CII]}=31$\,$M_\sun$\,$L_\sun^{-1}$ causes the molecular gas masses to be too high by $\gtrsim3\times$ following the direct non-detection of CO assuming Milky Way-like $\alpha_\mathrm{CO}$. 

Contrary to average conditions within the ISM of main sequence galaxies at high-$z$ \citep{Zanella2018,Madden2020,Dessauges-Zavadsky2020}, the [C\,{\sc ii}] streamer gas in SPT2349$-$56 -- with a (concordance) gas mass of $\sim$$0.89\times10^{10}$\,$M_\sun$ -- is 10$\times$ brighter in [C\,{\sc ii}] than expected for typical PDR conditions found within regular star-forming galaxies across the Universe.

\section{Discussion}\label{sec:discussion}

\subsection{Origin of the streamer gas}
We detected galaxy-sized, coherently flowing gas streamers fragmenting into kpc-sized clumps as traced with [C\,{\sc ii}]158$\mu$m emission, around a gas-rich galaxy triplet in the core of a protocluster at $z=4.3$. For a typical mass-to-light ratio, the dust continuum emission is faint while the [C\,{\sc ii}] line intensity is brighter than expected. The low velocity dispersions of $\sigma_{v,\mathrm{[CII]}}>31$\,km\,s$^{-1}$ is indicative of turbulence. Importantly, the morpho-kinematics of the streamers mirrors the gas motion of the three connected SMGs even out to $>25$\,kpc, evocative of a past ejection event.

Although nuclear outflows, entraining $M_\mathrm{gas}\approx10^{10}$\,$M_\sun$ of molecular gas around massive, high$-z$ galaxies, are found by \citet{Diaz-Santos2018}, \citet{Venemans2020}, \citet{Cicone2021}, or \citet{Meyer2022}, and a powerful radio source exists towards SMG `C' \citep{Chapman2024}, the overall morpho-kinematics of the [C\,{\sc ii}] streamers does not favor a nuclear outflow-dominated scenario akin to the above examples. 
Considering the current lack of high-resolution, outflow-sensitive tracers with rest-frame UV/optical instruments, and the remarkable space density of massive galaxies in this system, we will focus on tidal interactions as the main origin scenario for the streamers.

\subsubsection{Tidal stripping}\label{sec:tidal}

Galaxy mergers are a key element driving galaxy evolution \citep{Toomre1972,Toomre1977,Elmegreen2007,Duc2013}. Therefore, we consider whether the [C\,{\sc ii}] features are likely to be tidal streams of gas ejected by the `B'-`C'-`G' merger. The key points in favor of this scenario are the smooth velocity gradient across the streams and their connection in position-velocity space with the ULIRG triplet galaxies. One can estimate the survival of gas clumps in streams by comparing the tidal shearing force with the force of self-gravity as the morphology of the streamer clumps evolves in the tidal field of the protocluster core.

The tidal force $F_\mathrm{tidal}$\footnote{$F_\mathrm{tidal}=-R_\mathrm{cl}\frac{\mathrm{d}}{\mathrm{d}D}\left(\frac{G M_{<D}}{D^2}\right)=R_\mathrm{cl}\left(\frac{2GM_{<D}}{D^3}\right)$.} is acting at the surface of a clump with radius $R_\mathrm{cl}$ and as a function of distance $D$ to the barycenter with enclosed mass $M_{<D}$. 
At a distance $D$, smooth streamers are expected to develop out of stripped low-density clumps, i.e. self gravity is weaker than the external tidal force, forming tidal streams. 

The critical tidal mass of a clump $M_\mathrm{cl}$ given by $M_\mathrm{gas}>M_\mathrm{cl}=2M_{<D}({R_\mathrm{cl}}/{D})^3$ is a criterion for stabilizing self-gravity. $M_{<D}=1.43\times10^{11}$\,$M_\sun$ is assumed for the unresolved mass of SMG `C' \citep{Hill2022}. A value of $M_\mathrm{cl}/M_\mathrm{gas}<1$ indicates dominant clump self-gravity over tidal destruction. For parameters in Tab.~\ref{tab:results}, and $R_\mathrm{cl}\equiv R_\mathrm{spine}$ from Tab.~\ref{tab:results_continuum} we find upper limits for the tidal clump mass quotients of $M_\mathrm{cl}/M_\mathrm{gas}\lesssim 77$, $\lesssim 5$, $\lesssim 59$, and $\lesssim 8$ for `S1', `S3', `CS1', and `CS3'\footnote{For the counter-streamer sources, we assume the same central mass, $M_{<D}=1.43\times10^{11}$\,$M_\sun$, but substitute $D$ with the projected distance to `B' instead of `C' (see also Sec.~\ref{sec:gradients}).}, respectively. By correcting $D=D_\perp/\cos{(i=60^\circ)}$ for local spine inclination, more realistic stability quotients of $M_\mathrm{cl}/M_\mathrm{gas}\approx10$, $\approx0.6$, $\approx7.4$, and $\approx1.0$ are obtained.

Although the uncertainties grow quickly, the tidal stability criterion does predict critical stability for clump `S3’ for any spine inclinations larger than a low $i\approx30^\circ$. Segments `S1', `S6', `S7', `CS1', and `CS2' appear tidally unstable even after accounting for an inclination and thus larger 3-D distance to the central galaxies. 
However, the remaining IDs `S2'--`S5' give $M_\mathrm{cl}/M_\mathrm{gas}\lesssim2$, lending credence to a clump density structure expanding in the local tidal field. Importantly, \citet{Bournaud2011} predicted strikingly similar tidal clump structures to occur during high-$z$ mergers with high gas fractions. 

Another key characteristic of the [C\,{\sc ii}] streamers are smooth, coherently connected velocity gradients. 
To explore the idea of re-accreting tidal arms, we assume free falling gas clouds with velocity gradients $\nabla v_\mathrm{free}$ naturally forming as a function of radial distance $R$ and enclosed mass $M_{<R}$ denoted with $\nabla v_\mathrm{free}$\footnote{$\nabla v_\mathrm{free}\equiv-\frac{\mathrm{d}}{\mathrm{d}R}\sqrt{\frac{2GM_{<R}}{R}}=\sqrt{\frac{GM_{<R}}{2R^3}}.$}.
For the enclosed mass as above and a projected distance between `C' and `S2' of $R=10\,\mathrm{kpc}$, representative for the strait part $\nabla\mathrm{(S1-S5)}$, we obtain a free-falling gradient of $\nabla v_\mathrm{free}=17$\,km\,s$^{-1}$\,kpc$^{-1}$. Without any geometric corrections, a velocity gradient of $|\nabla v_\perp|=(235 - 118)\mathrm{km\,s}^{-1}/(6.2 - 13.2)\mathrm{kpc}=16.7$\,km\,s$^{-1}$\,kpc$^{-1}$ between `S1' and `S2' is found. Thus, free falling streamers seem to successfully explain the observed gradient among `S1'--`S2'. 

Again, assuming a 3-D disk-like rotation, with a constant phase angle $\phi=90^\circ$ and an inclination angle $i$ the equation above is then corrected by a factor of $\sin{(i)}\cos^{3/2}{(i)}$ to compare the projected gradients $\nabla v_{\mathrm{free},\perp}$ to the observed values. The extremum at $i=39.2^\circ$ yields $\nabla v_{\mathrm{free},\perp}\lesssim10.7$\,km\,s$^{-1}$\,kpc$^{-1}$.

Similarly, for the longer part of the streamer $\nabla\mathrm{(S3-S7)}$, $\nabla v_{\mathrm{free},\perp}\approx3.6$\,km\,s$^{-1}$\,kpc$^{-1}$ is observed. Given the observed velocity gradient is steeper by a factor of $\sim$2 (see Sec.~\ref{sec:gradients}), there is either a much larger mass aggregate close to the central SMGs, or radial re-accretion is not the correct mechanism\footnote{The toy model for $\nabla v_{\mathrm{free},\perp}$ assumes $\phi=90^\circ$ (for a maximal $\Delta v_\mathrm{[CII]}$). If we were to see the cloud at smoothly varying different phase angles, $\nabla v_{\mathrm{free},\perp}$ would drop, as only a smaller LOS velocity component is accessible in projection to the observer. $\nabla v_{\mathrm{free},\perp}$ is therefore treated as a strict upper limit.}.

Alternatively, it is possible that the longest part of the streamer is still expanding in 3-D, away from the initial ejection point, exerting compressive tidal forces on clumps `S3'--`S7' \citep{Renaud2008}. From this perspective, a substantial component of the 3-D velocity vectors are not pointing along our LOS but inherits the tangential velocity component from the disk. 
The undulating shape in the PV-diagram (see Fig.~\ref{fig:pv}) might be a feature of the projected variation in phase angle, modulating the observed velocity component of the streamers in sinusoidal fashion.

\subsubsection{Ram pressure stripping}\label{sec:rps} 

Ram pressure stripping (RPS) is an important mechanism to remove gas from a galaxy falling through a galaxy cluster \citep{Gunn1972,Boselli2022}. The combination of a high relative velocity between ISM gas and the hot medium filling the volume of the cluster can explain one-sided tails of material being pushed out of a galaxy \citep{Fumagalli2014}. 

In order to estimate the role of RPS in the creation of the [C\,{\sc ii}] streamers, a constant ambient density $\rho_\mathrm{ICM}$ is assumed, leading to $P_\mathrm{ram}=\rho_\mathrm{ICM}\Delta v^2$ from the surrounding proto-ICM. For instantaneous stripping to occur\footnote{The criterion $a(R)\Sigma_\mathrm{ISM}(R)=\frac{\sigma_\mathrm{[CII]}^2}{R}\Sigma_\mathrm{ISM}<\rho_\mathrm{ICM}\Delta v^2$ predicts ram pressure stripping to be effective at a critical radius $R$. The vertical gravitational acceleration is given by $a(R)\equiv a_\mathrm{disk}(R)=GM/R^2=GM_\mathrm{dyn}/(2 R_{1/2}^2)$.}, the pressure $P_\mathrm{ram}$ exerted on the galaxy while it moves through the ICM with relative velocity $\Delta v$ has to overcome the gravitational restoring force per unit area in the galactic disk $=a(R)\Sigma_\mathrm{ISM}(R)$.

Accordingly, a particle density of the ambient ionized medium of $n_\mathrm{ICM}=\rho_\mathrm{ICM}\mu m_\mathrm{H}=0.060(0.13)$\,cm$^{-3}$ (with $\mu=1$ as the mean molecular weight and $m_\mathrm{H}$ as the proton mass) is required for clump `S1'(`S1-peak'); for $\Delta v=500$\,km\,s$^{-1}$, we find $n_\mathrm{ICM}=0.013(0.038)$\,cm$^{-3}$.

The ICM density required to strip the gaseous outskirts of `S1' do vastly exceed the density at pressure balance of a typical hot ionized medium, i.e. $n_\mathrm{ICM}\gg n_\mathrm{HIM}=0.0035$\,cm$^{-3}$ \citep{McKee1977}. Furthermore, the densities for effectively removing material from the disk of ULIRG `G' -- possibly creating the western streamer -- are even higher at $n_\mathrm{ICM}=42$\,cm$^{-3}$ for $\Delta v=500$\,km\,s$^{-1}$. Therefore, RPS is not expected to remove material from galaxy `G' either (nor the more massive `B' and `C') as the densities required exceed the average densities of the cold neutral medium in a galaxy's ISM \citep{McKee1977}. Only a head-on collision scenario with a gas-rich disk can realistically strip cold clouds out of galaxy `G' \citep[cf.][]{Peterson2018,Yeager2020,Fadda2023}.

Considering the extended gas component reported by \citet{Zhou2025} with a total mass of $\sim2\times10^{11}\,M_\sun$ (for $\alpha_\mathrm{CO}=1$\,$M_\sun$\,K$^{-1}$\,km$^{-1}$\,s\,pc$^{-2}$) within a sphere of constant density with a radius equal to the maximum recoverable scale $\mathrm{MRS}=41.4$\,kpc of our ALMA observations, we find $n_\mathrm{ICM}=0.027$\,cm$^{-3}$. An encounter with this diffuse component could strip `S1', but only at high velocities ($\Delta v>500$\,km\,s$^{-1}$). 
Overall, while RPS might help to enhance tidal stripping, we do not expect it to be the main gas removal mechanism especially when compared to the tidal forces acting on the galaxies.

\subsection{Can molecular shocks explain the [CII] excess?}\label{sec:shocks}

\begin{figure*}[ht!]
\centering
\epsscale{1.0}
\includegraphics[height=0.31\textheight]{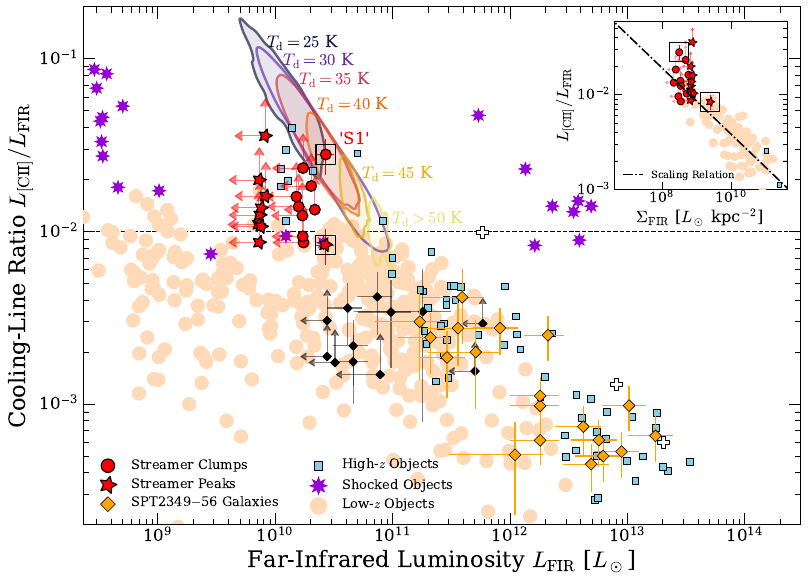}
\includegraphics[height=0.31\textheight]{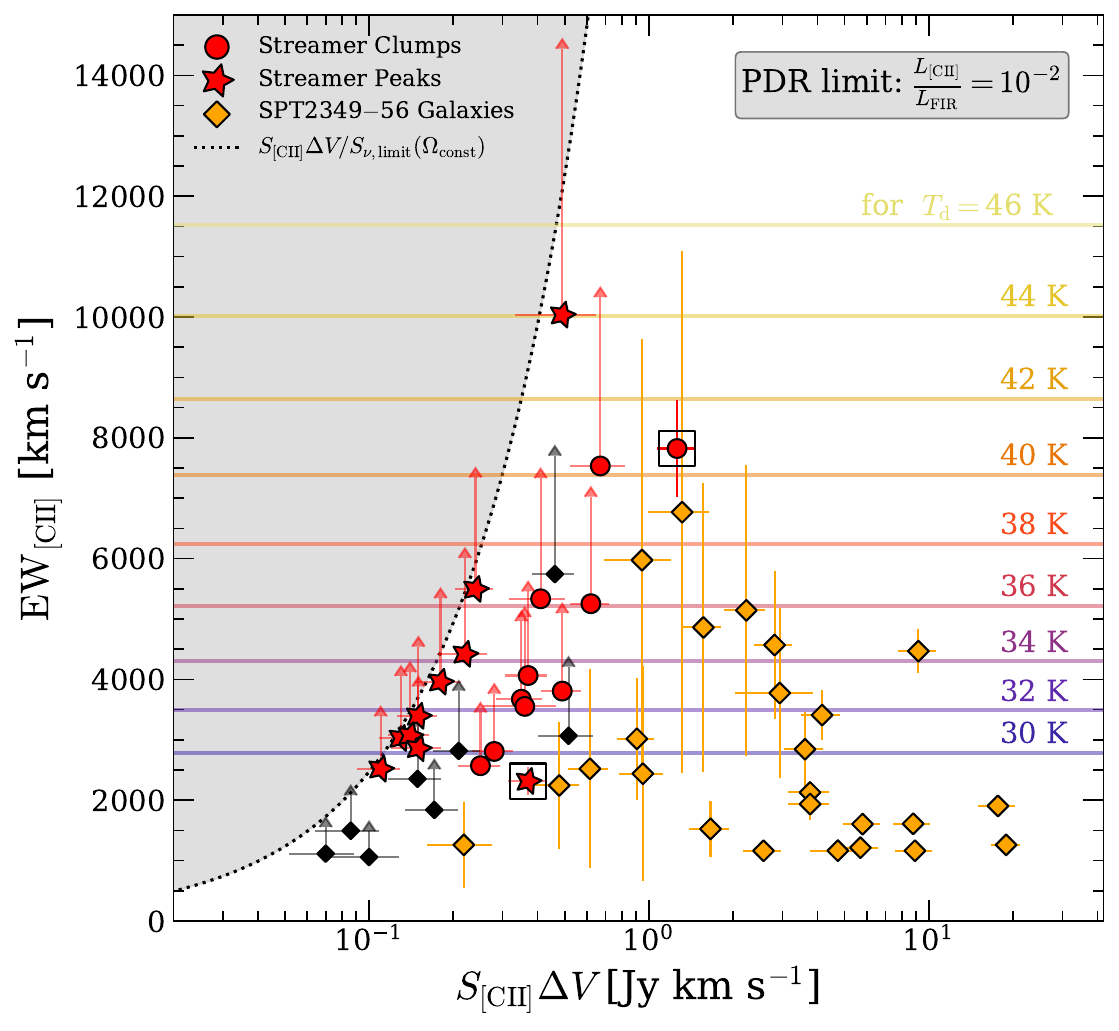}
\caption{
Left panel: Red circles (stars) represent the SPT2349$-$56 streamer clumps at $T_\mathrm{d}=30$\,K showing a [C\,{\sc ii}] ``excess''. \texttt{Model-0} dust temperatures and corresponding $L_\mathrm{FIR}$ (see Tab.~\ref{tab:models}) are indicated as blue ($T_\mathrm{d}=25$\,K) to yellow ($T_\mathrm{d}>50$\,K) contours. 
Local galaxies \citep{Brauher2008}, resolved gas clouds \citep{Fadda2021,Fadda2023} and interacting galaxies \citep{GOALS2018} are shown as peach circles. DSFGs in SPT2349$-$56 (yellow diamonds; \citealt{Hill2020}) and new sources (black diamonds) follow the high-$z$ ``deficit'', also seen in SMGs (\citealt{Gullberg2015, Carniani2018, Schaerer2020}; blue squares). Protocluster objects \citep{DeBreuck2022,Umehata2017} are shown as white symbols. Shocked clouds in \citet{Brisbin2015} PAH-bright galaxies, in Stephan's Quintet \citep{Appleton2013,Appleton2023}, and Taffy bridge sources \citep{Peterson2018} are shown as purple asterisks. UV-bright galaxies (\citealt{Capak2015}; blue squares) overlap in [C\,{\sc ii}]/FIR with the streamer clumps. Inlet shows the tight relation (\citealt{Diaz-Santos2013}; dash-dotted line) for [CII]/FIR versus far-infrared surface density $\Sigma_\mathrm{FIR}$ adapted from \citet{Spilker2016}. Right panel: [C\,{\sc ii}] equivalent width for SPT2349$-$56 streamer clumps and peaks, \citet{Hill2020} SMGs, and new candidates (black diamonds). PDR limit curves (horizontal lines) indicate a threshold for ``normal'' gas excitation ($\mathrm{[CII]/FIR}<1$\%). Our 3$\sigma_{850}$ upper limit criterion permits source detection outside the gray shaded region. `S1' (boxed circle) has the highest $\mathrm{EW}_\mathrm{[CII]}$ of any continuum detected source in the core. In the absence of shocks, warm dust conditions of $T_\mathrm{d}>40$\,K are required to explain $\mathrm{[CII/FIR}>1\%$ or the high $\mathrm{EW}_\mathrm{[CII]}$.
\label{fig:excess}}
\end{figure*}

In Fig.~\ref{fig:excess}, we show the results for the cooling line ratio $L_\mathrm{[CII]}/L_\mathrm{FIR}$ (see Tab.~\ref{tab:results_continuum}) for the [C\,{\sc ii}] streamer sources and various control samples. A [C\,{\sc ii}]/FIR ``deficit'' for FIR-luminous galaxies \citep{Malhotra2001} and the recently established [C\,{\sc ii}]/FIR ``excess'' at $L_\mathrm{[CII]}/L_\mathrm{FIR}>1$\% \citep{Appleton2013,Alatalo2014,Peterson2018,Posses2024} are immediately evident.

Our streamer clumps are found just at or even well into the ``excess'' regime, fully reversing the [C\,{\sc ii}] ``deficit'' trend for SMGs as they are completely isolated from any galaxy. For source `S1', with an assumed dust temperature of $T_\mathrm{d}=30$\,K and $\lambda_0=100$\,$\mu$m (see Sec.~\ref{sec:mass}), we find $\mathrm{[CII]/FIR}=2.8\pm0.5\times10^{-2}$. To illustrate the large uncertainty in $L_\mathrm{FIR}$, the full distribution of model posteriors for `S1' are included by utilizing \texttt{Model-0} (cf. Tab.~\ref{tab:models}). In Fig.~\ref{fig:excess}, this is shown by contours for $T_\mathrm{d}\in[25,30,35,40,45,>50]$\,K posteriors, marginalized over $\lambda_0$. The coldest $T_\mathrm{d}=25$\,K predicts a [C\,{\sc ii}] excess at a strict $>2\%$ level for `S1', regardless of $\lambda_0$. Warmer dust ($T_\mathrm{d}\geq 40$\,K) is compatible with the upper limit for a PDR origin of the [C\,{\sc ii}]/FIR ratio in `S1' but requires an extreme equivalent width $\mathrm{EW}_\mathrm{[CII]}=S_\mathrm{[CII]}\Delta V/S_{850}\gtrsim7500$\,km\,s$^{-1}$, as shown on the right panel in Fig.~\ref{fig:excess}. The model curves are upper limits for gas heated by PDRs ($\mathrm{[CII]/FIR}\ll1$\%) reflecting the most extreme kpc-scale galaxy properties \citep{Brauher2008,GOALS2018}. Despite the high dust temperatures required to produce the observed $\mathrm{EW}_\mathrm{[CII]}$, `S1' is expected to host even colder dust than the protocluster SMGs, due to the less effective radiative heating from young stars as suggested by the low SFR surface density (see also the inlet in Fig.~\ref{fig:excess}).

The high $\mathrm{EW}_\mathrm{[CII]}$ and $L_\mathrm{[CII]}/L_\mathrm{FIR}=2-3\times10^{-2}$ ``excess'' suggests that the [C\,{\sc ii}] streamers are more similar to collisional features between colliding galaxies, containing warm and only mildly photo-ionized gas, as found, e.g., in the Stephan's Quintet \citep{Appleton2013} and the Taffy galaxies \citep{Peterson2018}. 
This phenomenon is explained by galaxy-scale molecular shocks that collisionally excite the [C\,{\sc ii}]158$\mu$m transition via a supersonic turbulent energy cascade. 

Measurements of [NII]205$\mu$m in SPT2349$-$56 (D.~Zhou et al. in prep.) yielded a non-detection of [NII]205$\mu$m line emission\footnote{Equivalent to $S_\mathrm{[NII]}\Delta V<15.2$\,mJy\,km\,s$^{-1}$ (3$\sigma$) and integrated over a $\mathrm{FWHM}_\mathrm{[CII]}=95$\,km\,s$^{-1}$ at the location of `S3'.}, equivalent to $L_\mathrm{[CII]}/L_\mathrm{[NII]}\gtrsim53$ and thus well above the typical ISM-average value of SMGs \citep{Pavesi2016, Cunningham2020}. Ionized gas from HII regions, therefore does not seem to significantly contribute to the [C\,{\sc ii}] flux, further disfavoring PDR conditions inside the streamers.

Instead, the large surface intensity of [C\,{\sc ii}] in colliding galaxies originates from a warm, pressurized molecular medium with $n\approx10^3$\,cm$^{-3}$ and $T_\mathrm{gas}\approx160$\,K. To test whether a similar process might power the bright [C\,{\sc ii}] emission in SPT2349$-$56, we compare model surface intensities $I_\mathrm{[CII]}$ from the cold neutral and warm molecular medium to the observed 'S3-peak' surface intensity $I_\mathrm{[CII]}={(1+z)^4}/{\Omega_{\mathrm{bm}}}\int_{\nu} S_\mathrm{[CII]}\mathrm{d}\nu \simeq4.18\times10^{-11}$\,W\,cm$^{-2}$\,sr$^{-1}$. 
Hydrogen atoms collisionally excite the [C\,{\sc ii}]158$\mu$m transition in a cold neutral medium with $n_\mathrm{gas}=100$\,cm$^{-3}$ and $T_\mathrm{kin}=100$\,K, while electrons are negligible collision partners due the low ionization fraction \citep{McKee1977}. Following the analysis in \citet{Peterson2018}, a theoretical [C\,{\sc ii}] surface intensity\footnote{$I_\mathrm{[CII]} = \frac{h\nu_0A}{4\pi}\left[\frac{2e^{-91/T_\mathrm{kin}}}{1+2e^{-91/T_\mathrm{kin}} + n_\mathrm{cr}/n_\mathrm{gas} }\right]X_{C^{+}}N_\mathrm{gas}\Phi_\mathrm{bm}$ with solar carbon abundance \citep{Asplund2009}, C$^+$/C fraction of unity, HI column density of $N_\mathrm{gas}=N_\mathrm{HI}=3.5\times10^{21}$\,cm$^{-2}$, 36\% He fraction, and a critical density $n_\mathrm{cr}\approx 3000$\,cm$^{-3}$ \citep{Goldsmith2012} is assumed.} is found with $I_\mathrm{[CII]}=5.41\times10^{-12}$\,W\,cm$^{-2}$\,sr$^{-1}$. This result falls short by a factor of $\sim$8 compared to the observed surface intensity.

If we instead consider a pressurized molecular medium, experiencing adiabatic heating and assuming similar properties as in the Stephan's Quintet filament, i.e. $n_\mathrm{gas}=1000$\,cm$^{-3}$ and $T_\mathrm{kin}=160$\,K \citep{Appleton2013}, we expect $I_\mathrm{[CII]}=1.83\times10^{-11}$\,W\,cm$^{-2}$\,sr$^{-1}$, or $\sim$2$\times$ less than observed. 

From the measured surface density $\Sigma^\mathrm{[CII]}_\mathrm{gas}=38$\,$M_\odot$\,pc$^{-2}$ at 'S3-peak' (see Tab.~\ref{tab:results}), a density of $n_\mathrm{gas}=2670$\,cm$^{-3}$ within a spherical volume ($R_\mathrm{bm}=1.89$\,kpc) and a molecular volume filling factor\footnote{Molecular volume filling factors are typically $\Phi_V<2\times10^{-3}$\citep[e.g.][]{Pon2012,Klitsch2023}. Our lower value is roughly consistent with $\Phi_A$ following $T_\mathrm{b,[CII]}$ in Sec.~\ref{sec:mass}.} 
of $\Phi_V=10^{-4}$ is obtained. This higher surface density predicts a surface intensity of $I_\mathrm{[CII]}=3.34\times10^{-11}$\,W\,cm$^{-2}$\,sr$^{-1}$ -- fully in agreement with our observations. For $n_\mathrm{gas}=2670$\,cm$^{-3}$ and $T_\mathrm{kin}=300$\,K, corresponding to a thermal pressure $P_\mathrm{th}/k_\mathrm{B}=8.0\times10^5$\,K\,cm$^{-3}$, the observed and theoretical expectations for $I_\mathrm{[CII]}$ are equal within measurement uncertainties. Importantly, kinetic energy dissipation from galaxy-scale shocks towards supersonic, small-scale turbulence, manifesting as velocity dispersions of $\sigma_v\gtrsim30$\,km\,s$^{-1}$ can sufficiently maintain shock ionization, keeping the ionized carbon fraction close to unity in the streamer clumps.

\subsection{Kinematic structure of a monolithic collapse}\label{sec:dynamical}

\begin{figure*}[ht!]
\epsscale{1.0}
\centering
\includegraphics[width=0.999\textwidth]{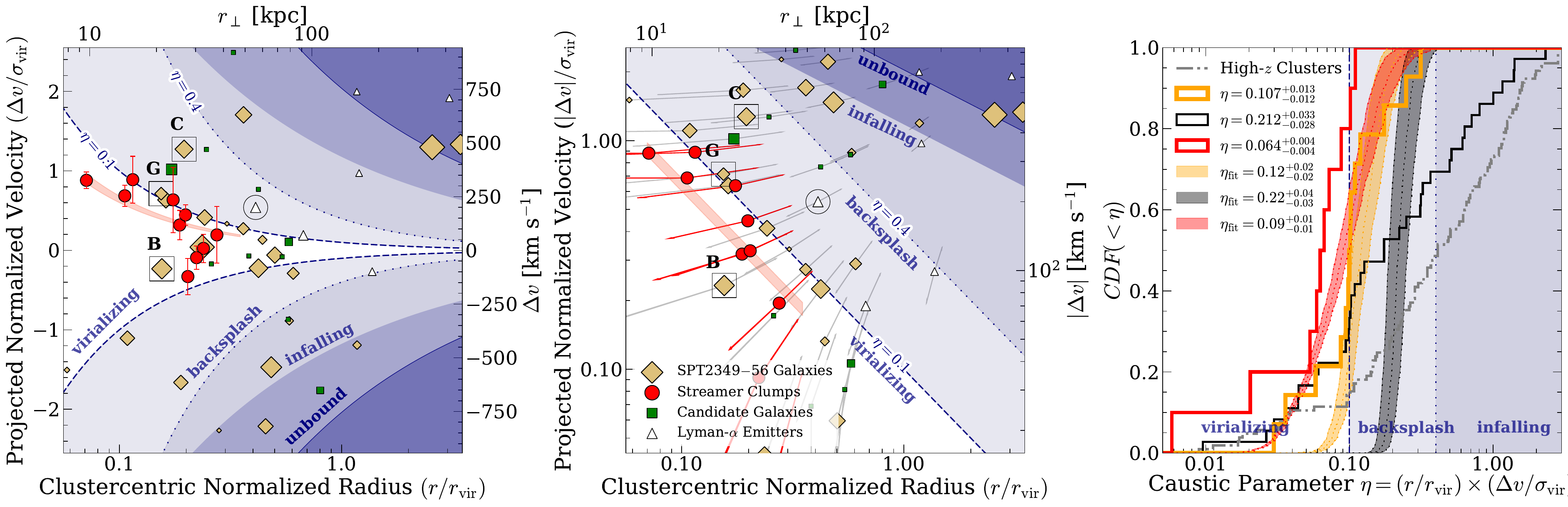}
\caption{Left panel: Red circles mark the phase-space distribution of streamer clumps with FWHM$_\mathrm{[CII]}$ as error bars. The line-of-sight velocities $\Delta v$ are normalized by the velocity dispersion $\sigma_{vir}=370$\,km\,s$^{-1}$ and projected radial distances $r$ by the virial radius $r_\mathrm{vir}=136$\,kpc. \citet{Hill2020} DSFGs are shown as yellow diamonds. Green squares are [C\,{\sc ii}] candidates, and white triangles denote LAEs \citep{Apostolovski2022}. Marker sizes scale with the line luminosity. The Lyman-$\alpha$ blob is marked by a circle. Center panel: The positional uncertainty around the 850$\mu$m common center position and systemic velocity is sampled and the magnitude of the propagated errors are indicated as straight lines in the [$-1\sigma_{r/r_\mathrm{vir}}$,$-1\sigma_{\Delta v/\sigma_\mathrm{vir}}$] and [$+1\sigma_{r/r_\mathrm{vir}}$,$+1\sigma_{\Delta v/\sigma_\mathrm{vir}}$] direction. Dark blue filled regions show the escape velocity $v_\mathrm{esc}=\sqrt{2GM_\mathrm{vir}/r}$ of the core as a function of radius with (solid line) and without de-projected $\Delta v$. Right panel: Comparison among caustic parameter $\eta$ (see Eq.~\ref{equ:eta}) cumulative density functions $CDF(<\eta)$ for source types (black curve for all sources in SPT2349$-$56 south, orange curve for DSFGs from \citet{Hill2020} with a $\eta<0.4$ cut, and red curve for the streamer clumps) to $z\sim1$ cluster galaxies (gray dashed curve; \citealt{Noble2013}). The shaded regions show the distributions of best fit $\eta$ (see main text) within one standard deviation. \label{fig:phase_space}}
\end{figure*}

The main formation scenario for early-type galaxies is via collisions between gas-rich disk galaxies leading to a dissipative merger event \citep{Toomre1977,Barnes1989,Bekki2001,Bournaud2011}. However, mergers of multiple galaxies in dense groups that occur on a shorter time scale than the individual molecular gas depletion times are less well understood \citep{Bekki2001} yet might be common at high-$z$ and responsible for extreme SFRs exceeding $>10^4$\,$M_\sun$\,yr$^{-1}$ \citep{Li2007}. These multiple merger events are at the heart of the debate about the hierarchical versus anti-hierarchical origin of giant elliptical galaxies \citep[see e.g.][]{DeLucia2006,DeLucia2007}. Traces of rare, short-lived multiple merger events should be visible in the velocity distribution of colliding galaxies and their lingering tidal debris. 

An important diagnostic tool for the gravitational collapse history is the halo phase space diagram \citep[see][]{Muzzin2014}. The location of galaxies in phase space encodes information about the infall or accretion time and general dynamical evolution of the galaxies during virialization of the system. Different studies \citep{Miller2018, Hill2020,Rennehan2020,Apostolovski2022} discuss SPT2349$-$56's the phase space configuration. Here we add a discussion of the possible connection between the synchronized star-formation episode and the dynamical state of the merging halo galaxies themselves \citep{Casey2016}.

Galaxies in the cores of clusters had sufficient time to partially virialize \citep{Mahajan2011} while structures at larger radii from the cluster's center preserve the initial orbital angular momentum as the projected, dimensionless observable, also referred to as caustic parameter
\begin{equation}\label{equ:eta}\eta\equiv(r/r_\mathrm{vir})\times (|v|/\sigma_\mathrm{vir})\end{equation} that characterizes the orbital history of a subsystem for a few crossing times. In this regard, accretion events of substructure will cause a characteristic, coherent phase space distribution with roughly constant $\eta$. Tidal torque acting along the trajectory will reduce $\eta$ with time as the galaxies' orbits virialize; meaning that the more recently the accretion occurred, the clearer the projected phase space signatures will be, appearing as ``streams'' and ``shells'' \citep{Romanowsky2012}.

For the back-splash population -- galaxies completing the second pericenter passage \citep{Pimbblet2011} at $0.1<\eta<0.4$ -- we might find well-defined differences in their cold gas properties and star-formation activity. However, mostly mild effects on high-$z$ galaxy properties are observed so far \citep{Noble2013,Muzzin2014,Wang2018,Williams2022,Perez-Martinez2022} despite enhanced susceptibility to tidal stripping \citep{Rhee2017}, galaxy collisions, and increased ram pressure \citep{Taylor2004,Yoon2017}. 

\subsubsection{Tidal streamers as beacons of halo assembly history}

Illustrated in Fig.~\ref{fig:phase_space}, a pattern emerges in the phase-space distribution of sources in the southern core of SPT2349$-$56. 
First, for $\eta<0.1$ and small radii, the halo is mostly unpopulated; only at larger distances, measured from the flux weighted 850$\mu$m center position defined in \citet{Hill2020} of 23$^\mathrm{h}$49$^\mathrm{m}$42.41$^\mathrm{s}$--56$^\circ$38$\arcmin$23.6$\arcsec$ (see Fig.~\ref{fig:map}; red asterisk) at $0.2<r/r_\mathrm{vir}<0.5$, four DSFGs 
are found with particularly low $\Delta v/\sigma_\mathrm{vir}$. Second, for galaxies at $20<r_\perp/\mathrm{kpc}<70$ but at enhanced normalized velocity, we find that roughly half the galaxies scatter around the $\eta\approx0.1$ caustic. Among them is central ULIRG `G' and strikingly all western streamer clumps at $\eta\lesssim0.1$. The counter-streamer seem to cross the projected zero-velocity boundary, but mirrors the $\eta$-distribution of the western streamer at negative velocity, indicative of large-scale streaming motion \citep[cf.][]{Romanowsky2012}. 

Furthermore, two LAEs and the Lyman-$\alpha$ blob \citep{Apostolovski2022} fall within the backsplash caustics $0.1<\eta<0.4$ but at strictly larger radii $r/r_\mathrm{vir}>0.3$. 
Also, the bright radio AGN `C' is found at $\eta=0.25$, while `B' is within the ``virializing'' domain $\eta=0.036$ and `G' is found at the backsplashing--virializing boundary ($\eta=0.11$). 
At $\gtrsim$300\,kpc, the bright northern SMGs `N1' and `N2' seem to be marginally bound to the southern core, as their velocities fall on (above) the escape velocity $v_\mathrm{esc}=\sqrt{2GM_\mathrm{vir}/r}$, with (without) a $\sqrt{3}$ velocity de-projection factor for the aforementioned halo mass. 

Despite a substantial positional scatter, many galaxies as well as the streamers seem to occupy similar caustics, hinting at a common infall history or dynamical origin. Quantitatively, we find median caustic parameters for $\eta<0.4$-selected SMGs and star-forming galaxies ($N=14$)\footnote{The median caustic parameter for the SMGs and star-forming galaxies on the back-splash caustic ($N=14$) is derived from: `B', `C', `D', `E', `F', `G', `I', `L', `M', `N', `C12', `C15', `C20', `C21'. SMGs `H', `J', and `K' are excluded due to their large offset in $\Delta v$. SMG `A' is removed from the list because it is very close to the systemic velocity with $\Delta v=15\pm4$ thus $\eta\approx0$.} of $\eta=0.107^{+0.013}_{-0.012}$ and for the streamer segments ($N=10$) $\eta=0.064^{+0.004}_{-0.004}$. Notably, eight galaxies (`D', `F', `G', `I', `L', `N', `C20', `C21') are just within $\pm$20\% of $\eta=0.1$. The median values between the back-splash SMGs and streamers are offset by $\Delta\eta\approx0.04$, but the distributions seem to follow unique caustics\footnote{For the backsplashing SMGs, we find a Pearson's coefficient of $R=-0.761$($p<0.001$) and Spearman's rank coefficient of $\rho=-0.714$($p<0.002$). For the streamer sources we find $R=-0.698$($p<0.012$) and Spearman-rank index of $\rho=-0.891$($p<0.001$) between $r/r_\mathrm{vir}$ and $\Delta v/\sigma_\mathrm{vir}$. Upper and lower limits reflect the 16$^\mathrm{th}$ to 84$^\mathrm{th}$ percentiles sampled from the measurement uncertainties.}
. 

On the contrary, when including all spectroscopically-confirmed galaxies in SPT2349$-$56 ($N=36$; excluding northern sources and [C\,{\sc ii}] streamer clumps, but including new [C\,{\sc ii}] candidates) we find $\eta=0.212^{+0.033}_{-0.028}$ with a Pearson's coefficient of $R=-0.094\,(p<0.292)$ and Spearman's rank coefficient of $\rho=-0.087\,(p<0.306)$. Thus negligible trends between $r/r_\mathrm{vir}$ and $\Delta v/\sigma_\mathrm{vir}$ are found. The cumulative distributions for these three populations are shown as red, orange, and black curves in the right panel of Fig.~\ref{fig:phase_space}. Interestingly, the SpARCS intermediate-$z$ cluster members (\citealt{Noble2013}; gray dashed CDF) show a comparable $\eta$ distribution, but without the excess around $0.1<\eta<0.2$.

An important caveat of this method is that within the $\Lambda$CDM paradigm, the assembling dark matter halo of the cluster would merge more rapidly than the gaseous components within the SMGs. 
This real dynamical center is not directly accessible to our observations, but can be inferred from the distribution of the satellites in phase space. 
For our calculations, we assume a halo mass of $M_\mathrm{h}= 9\pm 0.5\times10^{12}$\,$M_\sun$ \citep{Hill2020}, as measured from the kinematics of the 23 [C\,{\sc ii}]-brightest members of the southern core, with a characteristic viral radius $r_\mathrm{vir}=136$\,kpc and underlying velocity dispersion of $\sigma_\mathrm{vir}=370$\,km\,s$^{-1}$. 
In such a configuration, the orbital angular momenta of the SMGs would rapidly decay via dynamical friction and promote merging of the baryonic components as the components mix in phase space \citep[see e.g.][]{Mamon1992,Elmegreen2007}. 

To assess how robust our $\eta$ measurements are against the choice of the halo center position and systematic velocity offset, we assume a Gaussian uncertainty in the coordinates of $\sigma_r= 3$\arcsec ($\sim20$ kpc) and a $0.1\times \sigma_{vir}=37$\,km\,s$^{-1}$ error, roughly equal to the spectral bin width. The positional uncertainty corresponds to the distance between the 850$\mu$m center of mass measured with ALMA and APEX \citep{Miller2018}. The set of parameters [$\log{(r/r_\mathrm{vir})}$,\,$\log{|v/v_\mathrm{vir}|}$] is calculated and their joint distribution is then randomly sampled 1000 times by assuming offset origin positions in [$r$,\,$\Delta v$]. Each realization is fit by a linear regression curve with constant slope of $-1$ and a functional form of $\log{|\Delta v/v_\mathrm{vir}|}=\log{\eta_\mathrm{fit}}-\log{(r/r_\mathrm{vir})}$. 
Individual galaxy population medians and uncertainties are then sampled from the $\eta_\mathrm{fit}$ distributions.

With this approach, we find a caustic of $\eta_\mathrm{fit}=0.12^{+0.02}_{-0.02}$ (for back-splash galaxies),  $\eta_\mathrm{fit}=0.09^{+0.01}_{-0.01}$ (streamers), and $\eta_\mathrm{fit}=0.22^{+0.03}_{-0.03}$ (all southern core galaxies). 
Although the uncertainties erase our ability to measure trends within the central $\sim$20\,kpc, it seems clear that the [C\,{\sc ii}] streamers and SMGs are tightly coupled to the transition between virializing and backsplash caustics, even after considering major phase-space uncertainties. Moreover, $\sim$50\% of the realizations predict $0.1<\eta_\mathrm{fit}<0.4$ for the streamer segments, consistent with a back-splash caustic origin.

We interpret the tight coupling via a remarkable well populated caustic around $\eta\approx0.1$ between stripped gas and galaxies in the core of SPT2349$-$56 as the consequence of a ``monolithic collapse'' event. The isotropic dissipative collapse of primordial gas reminiscent of the mostly retired monolithic collapse scenario \citep{Eggen1962,Larson1974,Naab2009} provides an explanation for $z>4$ proto-BCG formation. From a massive collapsing gas cloud in free-fall, forming stars with high efficiency, a spheroid system could be assembled before the collapsing fragments fully dissipate their potential energy \citep{Larson1974}, preventing the gas from settling into a disk. The most massive galaxy members would tidally interacted first, losing angular momentum from dissipation via ejecting lower-$\eta$ tidal streamers i.e. the [C\,{\sc ii}] streamers and possibly the structure that is identified as the Lyman-$\alpha$ blob, while the larger halo population is currently backsplashing. 
Dynamical friction then leads to mass segregation \citep{Chandrasekhar1943} causing the massive galaxies 'B', 'C', and 'G' to rapidly dissipate their potential energy and to migrate towards the center of the halo, potentially triggering a merging instability \citep{Mamon1992}. In this frame-work, tidally ejected streamer gas roughly traces this mass segregation process, potentially ``freezing in'' the halo-scale back-splash trajectory.

\subsection{Comparing the tidal tails to numerical simulations}\label{sec:sim}

\begin{figure}[ht!]
\epsscale{1.0}
\centering
\includegraphics[width=0.47\textwidth]{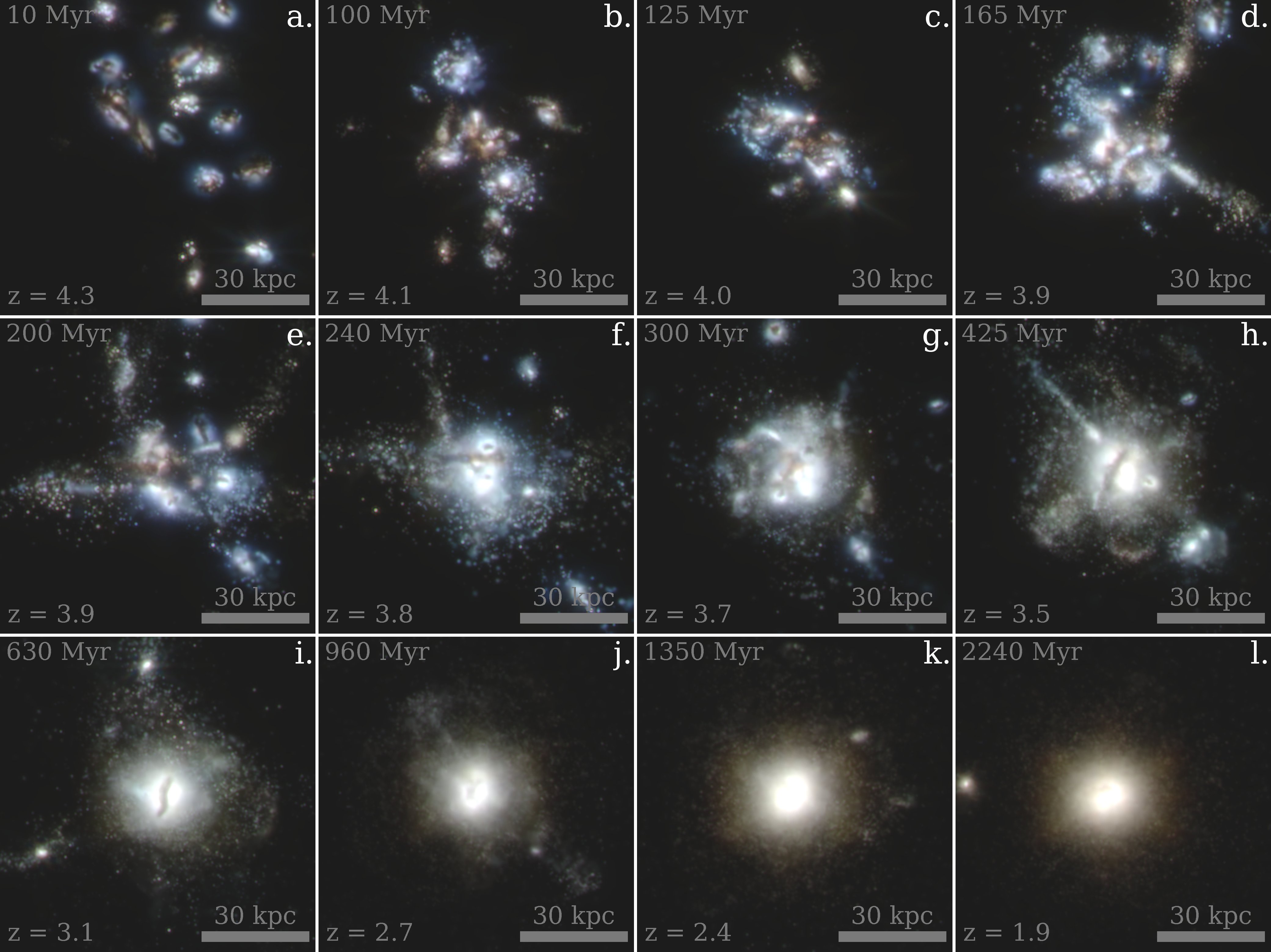}
\caption{ Synthetic \textit{JWST}/NIRCam snapshots tracking the time evolution of realization \texttt{real011} from the suite of SPT2349$-$56 analogs. The isolated \texttt{GIZMO} simulation was post-processed with radiative transfer code \texttt{SKIRT} and \texttt{STARBURST99} was used for the stellar population models. The mock RGB color composites are calculated for the filter bands F356W/F277W/F200W (as R/G/B). No evolving $K$-correction was applied to the filters, anchored to $z=4.3$. Note the numerous, 20\,kpc tidal steamers ejected around $\sim$200\,Myr. The size of the panels are 87\,kpc per side.\label{fig:sim}}
\end{figure}

Building upon the simulations in \citet{Rennehan2020}, an improved, non-cosmological $N$-body simulation suite, incorporating a total of 31 galaxies identified in SPT2349$-$56 (see \citealt{Hill2020} and Sec.~\ref{appendix:newgalaxies}) was conducted using a modified version of \texttt{GIZMO}\footnote{\url{http://www.tapir.caltech.edu/phopkins/Site/GIZMO.html}} \citep{Hopkins2015}, a hydrodynamics plus gravity program employing a mesh-free finite mass method to evolve the equation of motion. 
Initial conditions are sampled from realizations of 3-D radial distances ($R<95$\,kpc) and 3-D velocities drawn from the de-projected line-of-sight velocity distribution $v_\mathrm{3D}=\sqrt{3}\sigma_\mathrm{vir}$ per galaxy. Simulated gas masses were tied to the results from \citet{Hill2020}, \citet{Rotermund2021}, and this study (for new members).

Snapshots of simulation run \texttt{real011} are shown in Fig.~\ref{fig:sim}. It employs an average gas fraction of $f_\mathrm{gas}=M_\mathrm{gas}/(M_\ast+M_\mathrm{gas})=0.55$, a total stellar mass of $6\times10^{11}$\,$M_\sun$, a total gas mass of $7\times10^{11}$\,$M_\sun$, and constant total dark matter mass of $M_\mathrm{h}=2\times10^{13}$\,$M_\sun$. The mass resolution of baryonic particles is $10^6$\,$M_\sun$, and for dark matter particles it is $5\times 10^6$\,$M_\sun$; the simulation is progressed at time steps of 5\,Myr. Motivated by \citet{Venkateshwaran2024}, rotating disk initial conditions\footnote{\url{https://github.com/duncanmacintyre/galaxy-splicing}} were passed to \texttt{GIZMO}. Our sub-grid physics models are based on \citet{Dave2016} and \citet{Rennehan2019} but with implemented supermassive black hole growth as in \citet{Rennehan2020}. Feedback from AGNs was not included. 
The radiative transfer code \texttt{SKIRT} \citep{Camps2015} was employed and further post-processing used \texttt{STARTBURST99} \citep{Leitherer1999} to create synthetic \textit{JWST}/NIRCam images in filter bands F150W, F200W, and F277W at each time step \footnote{See \citet{Rennehan2020} for details on the post-processing.}. The projected intensity distributions were convolved with the corresponding NIRCam PSFs\footnote{\textit{JWST}/NIRCam point-spread functions for NIRCam filter bands F150W, F200W, and F277W from were taken from \url{https://stsci.app.box.com/v/jwst-simulated-psf-library/folder/174723156124}.} and sampled on a pixel grid at the native NIRCam resolution. The intensity maps per filter were then mapped to \texttt{RGB} color images and scaled with an area hyperbolic sine function to bring out fine details. 

In panel {\it a} of Fig.~\ref{fig:sim}, the initial simulation setup is shown and the first encounter of the galaxies occurs in {\it c}. 
As a result of the quasi spherical collapse, numerous giant tidal tails extend $D_\perp=20$--50\,kpc out from the center ({\it d} through {\it f}). Two types of tidal tails are identified: first, streams that connect to galaxies on highly eccentric orbits with large impact parameters on their second barycenter passage, are ejected like giant spiral arms. The dusty, star-forming ISM is carried away from the severely disrupted galaxies via these structures, visible in {\it e}. These giant arms possibly represent the stellar components to the observed [C\,{\sc ii}] streamers. Likewise, $N$-body simulations of gas-rich major mergers at $z=4$, show a similarly high fraction of 59\% in [C\,{\sc ii}] line luminosity emerging from tidal arms \citep{DiCesare2024} compared to our observations with $L_\mathrm{[CII]}^\mathrm{Ext.}/L_\mathrm{[CII]}^\mathrm{B-C-G}=34\pm$8\% in respect to $L_\mathrm{[CII]}$ from the merging galaxies. 

Second, we find tidal features characterized by radial accretion on oscillating orbits \citep[see also][]{Wetzel2011}. On the second barycenter passage, they stretch out to long, $\sim$30\,kpc straight streams, pointing away from the barycenter as the stripped galaxies plunge deeper into the central potential. Between snapshots {\it d} and {\it h}, numerous such tidal stripping events can be seen. 

Between $t=200$\,Myr and 500\,Myr, most of the stellar mass assembles {\it in situ} via collisions of several galaxies supplying cold gas. 
During the same time, a ring of dusty material becomes visible in absorption against the assembling early-type galaxy with radius $R\approx7$\,kpc. The ring slowly evaporates in the next 700\,Myr. 

In panels {\it h} and {\it i}, several satellites orbits or are oscillating radially around the main halo. By $t=1000$\,Myr, almost all of the satellites are completely disrupted by tidal stripping. 
One satellite galaxy is left, now orbiting the giant elliptical galaxy at $r\approx100$\,kpc until the end of the simulation run. 
Without rejuvination, the massive galaxy will likely become a quiescent elliptical galaxy with a stellar half-mass radius of $R_{50}\approx2.5$\,kpc and $M_\ast\approx8\times10^{11}$\,$M_\sun$. Hierarchical size growth via dissipationless dry mergers follows in the next few Gyr, consistent with previous simulations \citep{DeLucia2007,Ragone-Figueroa2018} and observation of high-$z$ BCGs \citep{Castignani2020}. By the end of the simulation, the tidal debris constitutes a faint but $\gtrsim$7\,kpc extended stellar component out to $\sim$65\,kpc and contains 25\% of the total mass.

Our simulations are expected to hold their validity until $t_\mathrm{cross}^\mathrm{N1}=r_\mathrm{N1}/\Delta v_\mathrm{N1}\approx0.71$\,Gyr or $z\approx3.0$. Thereafter, the northern component around HyLIRG `N1' \citep{Hill2020} might start to interact with the southern merger remnant. Even without this encounter, SPT2349$-$56 (south) might have morphologically grown into an object resembling MRC1138$-$262 \citep{Hatch2009,Emonts2016,Emonts2018} characterized by slightly lopsided multiple components surrounding the central stellar spheroid with visible dust lanes attenuating the $\sim$$10^{12}$\,$M_\sun$ stellar core.

\subsection{Survivability of the [CII] streamer clumps}\label{sec:survivability}

\subsubsection{Fragmentation of the molecular CGM}\label{sec:stability}

\begin{figure}[ht!]
\epsscale{1.0}
\includegraphics[width=0.46\textwidth]{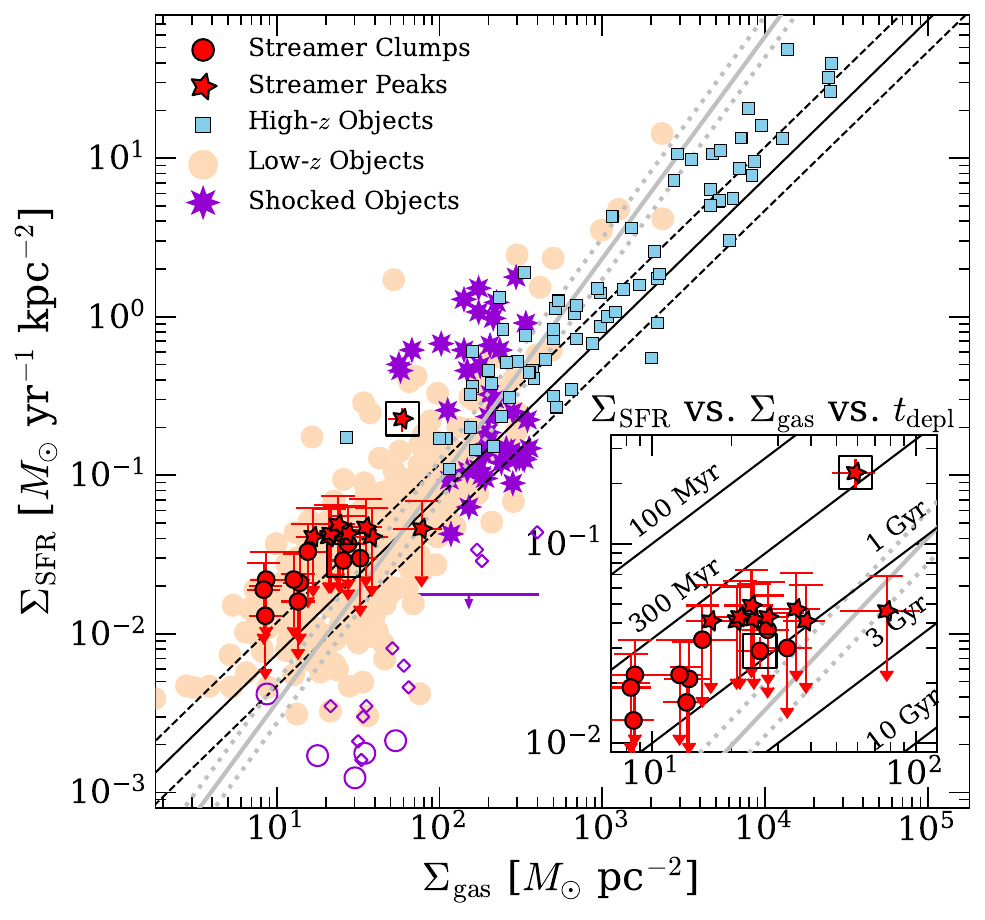}
\caption{The Schmidt-Kennicutt law suggests typical star-formation properties in the SPT2349$-$56 streamer clumps and peaks. Black boxes mark detected sources `S1' and `S1-peak', otherwise 3$\sigma$ upper limits are indicated as error bar caps while the circle or star symbols (for segments or peaks) are shown at the 2$\sigma$ limit. The Schmidt-Kennicutt law of nearby galaxies \citep{Kennicutt1998} is shown as the grey line, while the black trend line is the fit to molecular hydrogen in local spiral galaxies from \citet{Bigiel2008}. A low-$z$, non-resolved galaxy sample \citep{Saintonge2017} and high-$z$ galaxy samples \citep{Bethermin2020,Tacconi2013} are shown as peach circles and blue boxes, respectively. Sources from Stephan's Quintet \citep{Appleton2022} are plotted as purple stars. Kpc-scale resolved interacting galaxies and ram pressure-stripped clouds from \citet{Alatalo2014} and \citet{Moretti2020} (purple circles and diamonds), respectively, fall below the Schmidt-Kennicutt law. Inset shows lines of equal gas depletion time in black. 
}
\label{fig:KS}
\end{figure}

We now consider whether star formation could be occurring within the tidal streams of [C\,{\sc ii}] gas.
In Fig.~\ref{fig:KS}, the star-formation surface density $\Sigma_\mathrm{SFR}$ is compared to the molecular surface density $\Sigma_\mathrm{gas}$ for different galaxy samples \citep{Schmidt1959}. Our study extends the high-$z$ Schmidt-Kennicutt relation \citep[see e.g.][]{Tacconi2013,Decarli2019,Bethermin2023} down to $\Sigma_\mathrm{gas}<30$\,$M_\sun$\,pc$^{-2}$. The [C\,{\sc ii}] streamer clumps roughly occupy the same domain as local Universe star-forming galaxies \citep{Saintonge2017,Rosolowsky2021}, but slightly above the empirical \cite{Kennicutt1998} and \citet{Bigiel2008}. Employing equation Eq.~4 in \citet{Kennicutt1998}, we find that segment `S1' and `S3-peak' (3$\sigma$ upper limit of $\Sigma_\mathrm{SFR}$) fall on the Schmidt-Kennicutt relation within $<$2$\sigma$. `S1-peak' shows elevated $\Sigma_\mathrm{SFR}$ by a factor of $\sim$10 above the local relation. Ram pressure-stripped clouds \citep{Moretti2020} or clouds in tidal bridges \citep{Alatalo2014,Appleton2022} generally show about 10$\times$ lower $\Sigma_\mathrm{SFR}$ at the same gas surface density. Overall, our upper limit estimates for $\Sigma_\mathrm{SFR}$ are well within the limits for typical, local Universe spiral galaxy star-formation \citep{Bigiel2008}.  

Furthermore, all streamer sources show a cold gas depletion time of $t_\mathrm{depl}={\Sigma_\mathrm{gas}}/{\Sigma_\mathrm{SFR}}$ between 300\,Myr and $\lesssim$1\,Gyr. It can thus be expected that most of the ejected pre-heated gas reservoir will be dispersed on a dynamical time scale (see Sec.~\ref{sec:sim}) equal to or shorter than the cold gas depletion time of the system. 
This finding aligns well with shock heating as the source of the bright [C\,{\sc ii}] emission, where additional turbulent pressure might prevent efficient cooling of the gas in the ejected molecular clouds.

The ``turbulent Jeans mass'' $M_\mathrm{J} = {\sigma_\mathrm{turb}^4}/{(G^2\Sigma_\mathrm{gas})}$ with $\sigma_v=\sigma_\mathrm{turb}$, valid for kpc-scale segments of spiral arms in turbulent pressure equilibrium against self-gravity \citep{Toomre1964,Krumholz2005,Elmegreen2011}, predicts stability of gas with a $M/L$ of up to $\alpha_\mathrm{[CII],J} = \sqrt{\pi}{ R_\mathrm{spine}\sigma_\mathrm{turb}^2 }/{ (G L_\mathrm{[CII]} )}$. For $M_\mathrm{gas}>M_\mathrm{J}$ or $\alpha_\mathrm{[CII]}>\alpha_\mathrm{[CII],J}$, the streamer clumps become gravitationally unstable and start to fragment. Source ‘S3’\,('S3-peak') shows the lowest ratio between turbulent Jeans mass and gas mass at $M_\mathrm{J}/M_\mathrm{gas}=9.5$\,(3.1). This indicates that the local velocity dispersion is sufficient to hamper star-formation in this segment. Yet, if $\alpha_\mathrm{[CII]}$ was to increase above $\alpha_\mathrm{[CII],J}=5.2\,M_\sun/L_\sun$ (for `S3-peak'), turbulent pressure would not be strong enough to support the gas column density against collapse. We therefore conclude that the multi-scale turbulent energy cascade might have a negative impact on the small-scale star-formation process within the ejected clumps.

In addition, a different way to express the efficiency of star-formation is through the cloud free fall time per gas depletion time $\varepsilon_\mathrm{ff}=t_\mathrm{ff}/t_\mathrm{depl}={t_\mathrm{ff}\mathrm{SFR}_\mathrm{FIR}}/{M_\mathrm{gas}}={t_\mathrm{ff}\Sigma_\mathrm{SFR}}/{\Sigma_\mathrm{gas}}$ with $t_\mathrm{ff}=\sqrt{3\pi/32G\rho_0}$ denoting the free fall time of a spherical cloud with constant density $\rho_0=M_\mathrm{gas}R_\mathrm{spine}^{-3}/4\pi$ \citep{Krumholz2005}. With $t_\mathrm{ff}$ tabulated in Tab.~\ref{tab:results_continuum} we, again, find rather typical molecular cloud $\varepsilon_\mathrm{ff}$ of 15$\pm$2\,(2.8$\pm$1.7)\% for `S1'\,(`S1-peak') and $<$32\,($<$5.3)\% for 'S3'\,('S3-peak'). 
Local galaxy disks host $\varepsilon_\mathrm{ff}\approx5$\%, comparable to sources 'S1-peak' and `S3-peak', but lower than in high-$z$ spiral arms, where $\varepsilon_\mathrm{ff}$ reaches values of $30$--$50$\% \citep{Dessauges-Zavadsky2023}. To summarize, we find SFEs of $\sim$3\% to $<$21\% at the [C\,{\sc ii}] streamer intensity peaks, typical of local Universe to high-$z$ spiral arms, respectively.

\subsubsection{Chemical enrichment of the proto-ICM}\label{sec:enrichment}

Metals ejected through tidal streamers or expelled from subsequent supernovae would be locked up in the inner part of a pre-heated ICM, already by $z\approx3$. Therefore, the proto-ICM would inherit the elemental abundance pattern of cold gas in SMGs. Gas contributions from highly star-forming ISM conditions could relax the ``Metal Budget Conundrum'', an issue with a non-evolving, 40\%-solar Fe abundance \citep{Mantz2017} in the inner ICM of distant clusters that has not been fully resolved so far (\citealt{Mernier2016,Martin-Navarro2018,Blackwell2022,Blackwell2024}; but also see \citealt{Vogelsberger2018,Fukushima2023,Harada2023}).

The galaxy with the lowest gravitational binding energy among the ULIRG triplet is galaxy `G'. Although we can not accurately identify a single stripped galaxy in this three-body interaction (see Sec.~\ref{sec:tidal} and Sec.~\ref{sec:dynamical}), `G' is a likely candidate. Indeed, when comparing the total gas mass in the streamers ($8.9\pm0.4\times10^9\,M_\sun$) to the remaining molecular reservoir of `G' ($M_\mathrm{mol}=8\pm0.1\times10^9\,M_\sun$; \citet{Hill2020}), we find that a gas mass fraction of $\sim$53\% and $\sim$45\% without the counter-streamer might have been expelled from `G'.

To assess the plausibility of ICM metal enrichment via tidal ejection in SPT2349$-$56, we consider the average velocity gradient $\nabla v=14.5$\,km\,s$^{-1}$\,kpc$^{-1}$ along the `S1'--`S5' arc at $D_{\perp}=15.6$\,kpc within which we expect an enclosed gas mass of $M_\mathrm{gas}=44.1\times10^8$\,$M_\sun$. In combination, this yields a mass outflow rate of $\dot{M}_\mathrm{out}=M_\mathrm{gas}\nabla v\approx 65$\,$M_\sun$\,yr$^{-1}$. Therefore, the hypothetical mass loading factor for SMG `G' would be $\dot{M}_\mathrm{out}/\mathrm{SFR}_\mathrm{G}=0.38^{+0.16}_{-0.11}$ -- fully consistent with the aforementioned $\sim$50\%-stripped gas fraction of `G'. This value means that `G' has possibly lost roughly two thirds of gas mass to star-formation and one third to the build-up of a metal-rich CGM via tidal ejection. 

\citet{Spilker2020} found similar mass loading factors at a 30--50\% level in molecular outflows of lensed SMGs. Therefore, tidal stripping in protocluster cores could be as important for metal enrichment of the early ICM as stellar feedback driving molecular outflows \citep[see also][]{DiCesare2024}. Nevertheless, the role of AGN-driven outflows for the metal enrichment of the proto-ICM -- especially from the most massive protocluster members \citep{Chapman2024} -- remains to be established.

\section{Conclusion}\label{sec:conclusion}

We present new observations of the extended cold gas reservoir surrounding a triplet of ULIRGs in the protocluster core of SPT2349$-$56 at $z=4.303$ as traced by [C\,{\sc ii}]158$\mu$m emission. Utilizing combined ultra-deep ALMA Band-7 data, the structure is resolved into multiple thin arcs, fragmenting into massive, gaseous clumps. By characterizing the physical properties, origin, and possible future evolution of the clumps embedded in the giant [C\,{\sc ii}] streamers, we have establish a connection between the circumgalactic medium, the SMGs in the protocluster core, and the rapid assembly of a brightest cluster galaxy  with an expected stellar mass of $M_\ast\approx 10^{12}\,M_\sun$ by $z=3$. The main findings of this work are as follows:

\begin{itemize}

\item SPT2349$-$56 hosts a rare triplet of interacting, FIR-ultraluminous galaxies ($L_\mathrm{FIR}\approx10^{12-13}$\,$L_\sun$) in its core, connected by giant, coherent streamers of ionized carbon ions detected in [C\,{\sc ii}]158$\mu$m emission. A larger western ($l_\mathrm{3D}>50$\,kpc) and smaller eastern feature ($l_\mathrm{3D}>20$\,kpc) are found that loop back in both position and velocity to the ejection source, identified as a likely multiple major merger. Like ``beads on a string'' the clumpy tidal arms, reminiscent of local gaseous HI streams \citep{deBlok2018, Jones2019} or the CGM in high-$z$ quasars \citep{Diaz-Santos2018, Decarli2019}, fragment into ten clumps in total, each $\sim$5\,kpc in size, with low velocity dispersions in the western part of $\sigma_\mathrm{[CII]}=40$--$130$\,km\,s$^{-1}$, centered around $\Delta v_\mathrm{[CII]}=0$--$330$\,km\,s$^{-1}$.

\item The combined line luminosity of the [C\,{\sc ii}] streamers is $L_\mathrm{[CII]}=30.2\pm1.9\times10^8$\,$L_\sun$ corresponding to an intensity of $S_\mathrm{[CII]}\Delta V=5.06\pm 0.33$\,Jy\,km\,s$^{-1}$. Co-spatial with the [C\,{\sc ii}] surface brightness peak of clump `S1' is an unresolved dust continuum source with 850\,$\mu$m flux density of $S_\mathrm{850}=0.15\pm3$\,mJy. For typical cold dust conditions ($T_\mathrm{d}=30$\,K) we find $M_\mathrm{mol}=22.2\pm4.1\times10^8$\,$M_\sun$ of molecular gas in `S1', mapping to a low mass-to-light ratio of $\alpha_\mathrm{[CII]}=2.95\pm0.30$\,$M_\sun$\,$L_\sun^{-1}$. This factor is $\sim$10 times lower than typically found in $z\approx 4$ star-forming galaxies \citep{Zanella2018} hinting at limited predictive power of the [C\,{\sc ii}] line luminosity as a cold gas mass tracer for high-$z$ extended gas. Accordingly, the total molecular gas reservoir in the streamers is $M_\mathrm{mol}=8.9\pm0.7\times10^9\,\frac{\alpha_\mathrm{[CII]}}{2.95}$\,$M_\sun$.

\item For `S1', the boost of [C\,{\sc ii}] line emission over the dust continuum of $\mathrm{EW}_\mathrm{[CII]}=8200\pm1900$\,km\,s$^{-1}$ or $\mathrm{[CII]/FIR}=2.8\pm0.7\times10^{-2}$ ($T_\mathrm{d}=30$\,K) is best explained by molecular shocks redistributing mechanical energy via a turbulent energy cascade. A possible warm, pressurized molecular medium measured at the location of `S3-peak' with $n_\mathrm{gas}\approx3\times10^3$\,cm$^{-3}$, $T_\mathrm{gas}\approx300$\,K, and $P_\mathrm{th}/k_\mathrm{B}\approx10^6$\,K\,cm$^{-3}$, shows similar conditions as found in local Universe galaxy-scale shocks.

\item In phase-space, the streamer segments are observed to follow a narrow range around a caustic parameter $\eta$ i.e. curves of constant specific angular momentum of $\eta=(r/r_\mathrm{vir})\times(|v|/\sigma_\mathrm{vir})=0.064^{+0.004}_{-0.004}$. Fourteen galaxies are found to populate caustics of $\eta=0.107^{+0.013}_{-0.012}$, close to the demarcation between the back-splashing and virializing domain. This tight coupling between the streamers clumps and the protocluster galaxies point towards a common dynamical origin, potentially related to the halo virialization following the spherical collapse of the halo gas.

\item We find a SFE of $\varepsilon_\mathrm{ff}\approx3$\% to $<$21\% (3$\sigma$) at the location of the peak [C\,{\sc ii}] intensities of the streamer clumps, in line with turbulent Jeans masses that indicate kpc-scale cloud stability against fragmentation. Falling close to the Schmidt-Kennicutt relation, the [C\,{\sc ii}] streamer gas is likely to be more rapidly dynamically dispersed than it might be consumed by star-formation ($t_\mathrm{depl}\gtrsim300$\,Myr), thus assembling a multi-phase circumgalactic medium. Our $N$-body simulations show that within the next $\sim$300\,Myr, repeated merging of gas-rich galaxies with the proto-BCG might persistently strip galaxies of $\sim$30--50\% of their molecular gas, thus significantly contributing to the injection of metals into the proto-ICM.

\end{itemize}

The bright ionized carbon streamers are a striking signpost for massive galaxy formation in the distant Universe. Our deep ALMA observations of [C\,{\sc ii}]158$\mu$m line emission aid the characterization of the extended gas morpho-kinematics and thus provide new evidence for the monolithic dissipative collapse of a group of gas-rich galaxies at $z=4.3$. However, our estimates of the physical properties of the cold dust in the system -- and thus by proxy the molecular gas mass -- are so-far limited by the maximum recoverable scale and frequency coverage of our observations. High-frequency observation in the THz-range at $\lesssim$0.5$\arcsec$ are necessary to unambiguously probe the dust content of the SMGs and the [C\,{\sc ii}] streamers. 

\begin{acknowledgments}

N.S. gratefully acknowledges the Collaborative Research Center 1601 (SFB~1601 project C1) funded by the Deutsche Forschungsgemeinschaft (DFG, German Research Foundation) - 50070025.
MA is supported by FONDECYT grant number 1252054, and gratefully acknowledges support from ANID Basal Project FB210003 and ANID MILENIO NCN2024\_112.
This paper makes use of the following ALMA data: 
ADS/JAO.ALMA \#2016.0.00236.T, 
ADS/JAO.ALMA \#2017.1.00273.S, 
ADS/JAO.ALMA \#2018.1.00058.S,
ADS/JAO.ALMA \#2021.1.01063.S. ALMA is a partnership of ESO (representing its member states), NSF (USA) and NINS (Japan), together with NRC (Canada), NSTC and ASIAA (Taiwan), and KASI (Republic of Korea), in cooperation with the Republic of Chile. The Joint ALMA Observatory is operated by ESO, AUI/NRAO and NAOJ.


\end{acknowledgments}


\facilities{APEX LABOCA \citep{Siringo2009}, ALMA Band-7 \citep{Mahieu2012}, VLT MUSE \citep{Bacon2010}, \textit{HST}. }

\software{
\texttt{astropy} \citep{Astropy2022},
\texttt{CARTA} \citep{Comrie2021},
\texttt{CASA} \citep{McMullin2007,Petry2012},
\texttt{matplotlib} \citep{Hunter2007},
\texttt{MIRIAD} \citep{Sault1995},
\texttt{numpy} \citep{Harris2020},
\texttt{pvextractor} \citep{Ginsburg2016},
\texttt{scipy} \citep{Scipy2020}.
}



\newpage
\appendix

\section{Moment maps (zoom-ins)}\label{appendix:moments}

\begin{figure*}[ht!]
\centering
\epsscale{1.0}
\includegraphics[width=0.999\textwidth]{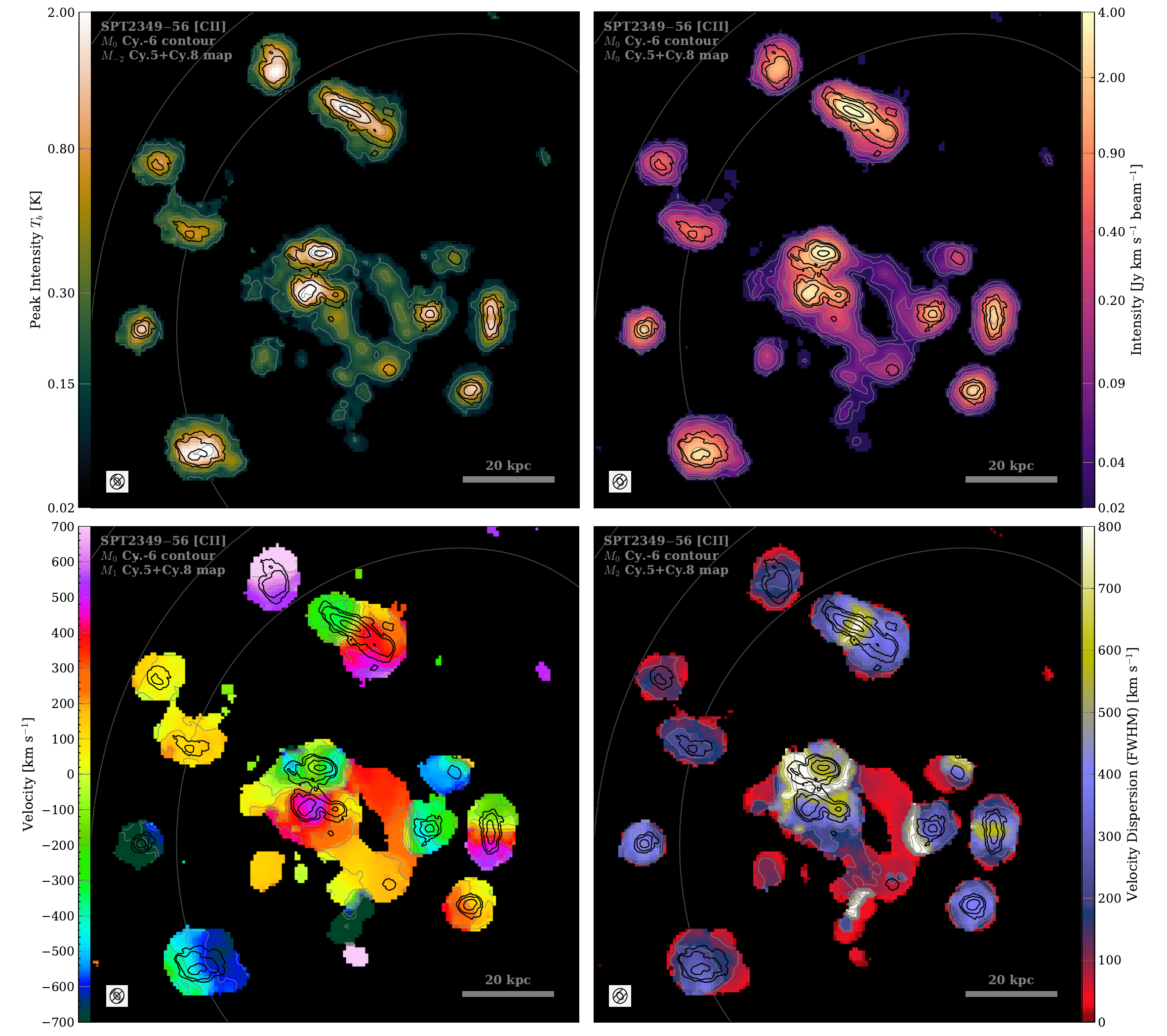}
\caption{Zoom-in of moment maps of the protocluster core SPT2349$-$56 in [C\,{\sc ii}]158$\mu$m line emission. $M_{-2}$ refers to the peak voxel intensity in brightness temperature units, converted from peak Jy\,beam$^{-1}$ using $T_\mathrm{b,[CII]}=(1+z)c^2 S_\mathrm{peak,[CII]}/(2 k_B\nu_{obs}^2\Omega_\mathrm{spine})$. Black contours trace Cycle 6-only [C\,{\sc ii}] emission at $\sim$0.2$\arcsec$ angular resolution with [0.03, 0.15, 0.6, 1.8, 3.6]\,Jy\,km\,s$^{-1}$\,beam$^{-1}$ intensity. Gray contours denote the primary beam at [$1/1.5$, $1/2$, $1/3$]$\times$ the phase center illumination.
\label{fig:momentszoom}}
\end{figure*}

Similar to the moment maps shown in Fig.~\ref{fig:map}, Fig.~\ref{fig:momentszoom} shows zoom-in maps without labeling of the central ULIRG triplet within $R=50$\,kpc. In addition, the peak intensity map ($M_{-2}$, following \texttt{MIRIAD} notation) is shown.

\newpage
\section{Dust continuum properties of the [CII] streamers}\label{appendix:continuum}
Tab.~\ref{tab:results_continuum} summarizes the dust continuum-derived results and highlights important quantities from the dust modified blackbody SED fitting to the continuum detection of streamer clump `S1'. See sections \ref{sec:mass} for details. We employ Eq.~\ref{equ:sobs} to obtain dust masses $M_\mathrm{d}$, assuming a gas-to-dust conversion factor $\delta_\mathrm{GDR}=100$ and Eq.~\ref{equ:fir} for far-infrared luminosities $L_\mathrm{FIR}$. Since `S1' is the only source detected at $\nu_{obs}=346.4$ GHz with $S_{850}=0.15\pm0.03$\,mJy, all other measurements are provided as 3$\sigma$ upper limits at the peak pixel position in the [C\,{\sc ii}] $M_0$ map. For a primary beam response of unity and a source size smaller than the synthesized beam size of $\Omega_\mathrm{bm}=0.233$\,arcsec$^2$, we find 3$\sigma_{850}=0.004$\,mJy\,beam$^{-1}$ for the combined ALMA data set.

\begin{deluxetable}{lcccccccccc}[!ht]
\tablenum{4}
\tablecolumns{10}
\tabletypesize{\scriptsize}
\tablecaption{Dust continuum flux densities and derived properties of the [C\,{\sc ii}] streamer sources in SPT2349--56.  }
\label{tab:results_continuum}
\tablehead{
\colhead{ID} & 
\colhead{$R_\mathrm{spine}$} & 
\colhead{$S_{850\mu\mathrm{m}}$} & 
\colhead{$L_\mathrm{FIR}$} & 
\colhead{SFR$_\mathrm{FIR}$} & 
\colhead{$\Sigma_\mathrm{SFR}$} & 
\colhead{$\mathrm{[CII]}/\mathrm{FIR}$} & 
\colhead{$t_\mathrm{ff}$} & 
\colhead{$\epsilon_\mathrm{ff}$} & 
\colhead{$M_\mathrm{d}$} & 
\colhead{$\alpha_\mathrm{[CII]}$} \\
  &
[kpc]  & 
[$\mu$Jy]  &
[$10^8L_\sun$]  & 
$\left[\frac{M_{\sun}}{\mathrm{yr}}\right]$  &
$\left[\frac{M_{\sun}}{\mathrm{yr\;kpc}^2}\right]$  &
[$10^{-2}$]&
[Myr]  &
[\%]  &
[$10^6M_\sun$]  &
$\left[\frac{M_{\sun}}{L_\sun}\right]$ 
}
\decimalcolnumbers
\startdata
S1(C16)  &  5.26  &  153$\pm$27  &  269$\pm$48  &  2.6$\pm$0.5  &  0.029$\pm$0.005  &  2.8$\pm$0.5  &  135$\pm$12  &  15$\pm$4  &  22.2$\pm$3.9  &  2.95$\pm$0.3\\
S2(C23)  &  3.88  &  $<$87    &  $<$158    &  $<$1.5    &  $<$0.032    &  $>$1.4  &  157$\pm$15    &  $<$36    &  $<$13.1    &  $<$5.91 \\
S3(C22)  &  5.10  &  $<$112    &  $<$202    &  $<$1.9    &  $<$0.024    & $>$1.8  &  183$\pm$17    &   $<$32    &  $<$16.7    &  $<$4.49 \\
S4  &  5.66  & $<$123	   &  $<$217    &  $<$2.1    &  $<$0.020    &  $>$1.3  &   242$\pm$23    & $<$58    &  $<$17.9    &  $<$6.15 \\
S5  &  4.01  &  $<$93    &  $<$174    &  $<$1.7    &  $<$0.033    &  $>$0.9  &  202$\pm$21    &  $<$77    &  $<$14.4    &  $<$9.74 \\
S6  &  4.31  &  $<$95    &  $<$171    &  $<$1.6    &  $<$0.028    &  $>$1.0  &  214$\pm$21    & $<$72    &  $<$14.1    &  $<$8.62 \\
S7  &  3.96  &  $<$91    &  $<$170    &  $<$1.6    &  $<$0.033    &  $>$1.2  &  166$\pm$17    &  $<$43    &  $<$14.1   &  $<$6.66 \\\tableline
CS1  &  3.40  &  $<$85    &  $<$172    &  $<$1.6    &  $<$0.045    &  $>$2.3  &   96$\pm$12    & $<$13    &  $<$14.2    &  $<$3.57 \\
CS2  &  2.89  &  $<$73    &  $<$151    &  $<$1.4   &  $<$0.055    &  $>$1.6  &  96$\pm$10    &  $<$19    &  $<$12.5    &  $<$5.11 \\
CS3  &  3.61  &  $<$96    &  $<$210    &  $<$2.0    &  $<$0.049    &  $>$1.0  &  143$\pm$24   &  $<$45    &  $<$17.4    &  $<$8.04 \\
\tableline
S1-peak  &  1.89 &  152$\pm$27  &  267$\pm$47  &  2.5$\pm$0.4  &  0.227$\pm$0.04  &  0.8$\pm$0.2  &  26$\pm$5  &  3$\pm$3  &  22.1$\pm$3.9  &  9.91$\pm$1.0\\
S2-peak  &  1.89  &  $<$42    &  $<$74    &  $<$0.7    &  $<$0.063    &  $>$1.2  &   40$\pm$8    &  $<$10    &  $<$6.1    &  $<$6.64 \\
S3-peak  &  1.89  &  $<$42    &  $<$73    &  $<$0.7    &  $<$0.062    & $>$2.0   &   32$\pm$6    &  $<$ 5    &  $<$6.0    &  $<$4.16 \\
S4-peak  &  1.89  &  $<$41    &  $<$72    &  $<$0.7    &  $<$0.061    &  $>$1.1  &   44$\pm$9    &  $<$13    &  $<$5.9    &  $<$7.49 \\
S5-peak  &  1.89  &  $<$43    &  $<$76    &  $<$0.7    &  $<$0.065    &  $>$1.1  &   43$\pm$9    &  $<$13    &  $<$6.3    & $<$7.75 \\
S6-peak  &  1.89  &  $<$42    &  $<$73    &  $<$0.7    &  $<$0.062    &  $>$0.9  &   49$\pm$10   &  $<$21    &  $<$6.0    &  $<$9.53 \\
S7-peak  &  1.89  &  $<$43    &  $<$76    &  $<$0.7    &  $<$0.064    &  $>$1.4  &   38$\pm$8    &  $<$10    &  $<$6.3    &  $<$6.03 \\\tableline
CS1-peak  &  1.89  &  $<$47    &  $<$82    &  $<$0.8    &  $<$0.069    &  $>$3.6  &   23$\pm$8    &  $<$2    &  $<$6.7    &  $<$2.30 \\
CS2-peak  &  1.89  &  $<$47    &  $<$83    &  $<$0.8    &  $<$0.071    &  $>$1.6  &   34$\pm$8    &  $<$7    &   $<$6.9    &  $<$5.17 \\
CS3-peak  &  1.89  &  $<$50    &  $<$88    &  $<$0.8    &  $<$0.074    &  $>$1.0  &   41$\pm$9    &  $<$12    &  $<$7.2    &  $<$8.03 \\
\tableline
$\int$\,Ext.  &  ---  &  ---  & 1894$\pm$188 & 18.0$\pm$1.8 & --- & --- & --- & ---  &  157$\pm$9  &  ---  \\
\tableline
\enddata
\tablecomments{
$^{(12)}$Source IDs from \citet{Hill2020} that are co-spatial with new sources are indicated in brackets. $^{(13)}$The flux densities in ${(14)}$ are measured for a spine size of $A_\mathrm{spine}=\pi R_\mathrm{spine}^2$. 
$^{(14)}$ $S_{850\mu\mathrm{m}}$ is the continuum flux density; upper limit values refer to the 3$\sigma$ dust continuum detection threshold and are not additionally scaled by 15\% systematic uncertainty. 
$^{(15)}$The far-infrared luminosities ($L_\mathrm{FIR}$) are calculated by numerically integrating the modified blackbody over 42--500 $\mu$m. 
$^{(18)}$[C\,{\sc ii}] over far-infrared cooling line ratio.
$^{(19)}$Cloud free-fall time, derive as $t_\mathrm{ff}=\sqrt{3\pi(32 G \rho_0)^{-1}}$ whereas the constant cloud density is approximated by $\rho_0=M^{\mathrm{[CII]}}_\mathrm{gas}(4\pi R_\mathrm{spine}^3/3)^{-1}$. 
$^{(20)}$SFE per free-fall time is evaluated by the equation $\epsilon_\mathrm{ff}=\mathrm{SFR}_\mathrm{FIR}(M^{\mathrm{[CII]}}_\mathrm{gas}/t_\mathrm{ff})^{-1}$. 
$^{(22)}$By assuming the gas-to-dust ratio of $\delta_\mathrm{GDR}=100$ we obtain the mass-to-light ratio $\alpha_\mathrm{[CII]}$ for `S1' and `S1-peak' and 3$\sigma$ upper limits for all other sources without additional calibration uncertainty.}
\end{deluxetable}

\newpage
\section{Position--velocity diagrams}\label{appendix:pvdiagrams}

\begin{figure}[!ht]
\epsscale{1.0}
\center
\includegraphics[height=0.35\textwidth]{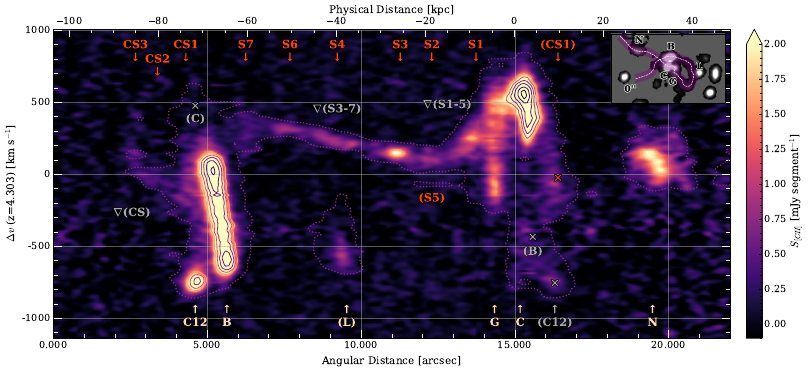}\\
\includegraphics[height=0.35\textwidth]{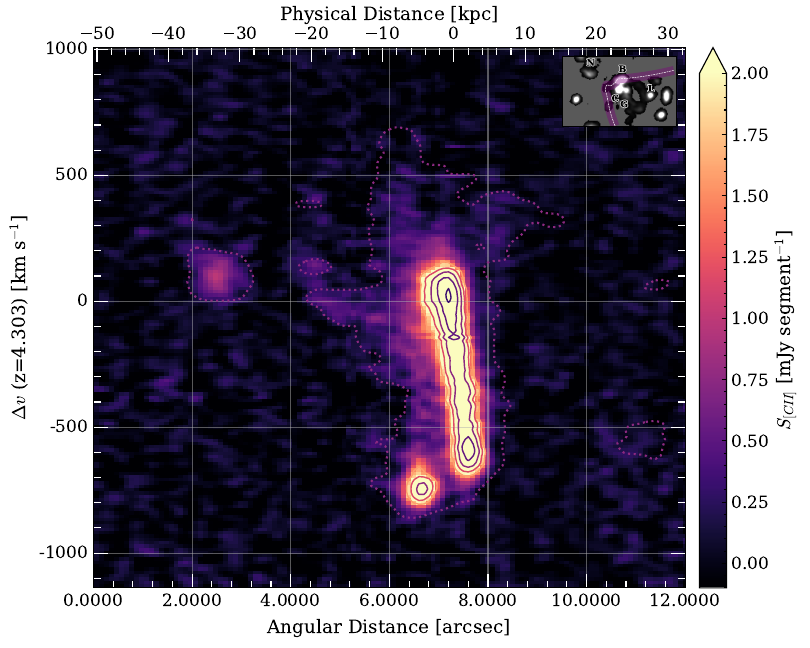}\hspace{-8.9mm}
\includegraphics[height=0.35\textwidth]{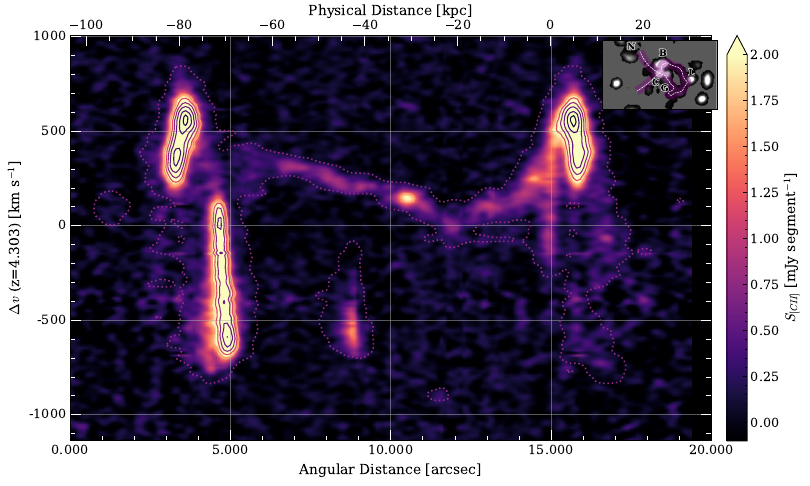}
\includegraphics[height=0.35\textwidth]{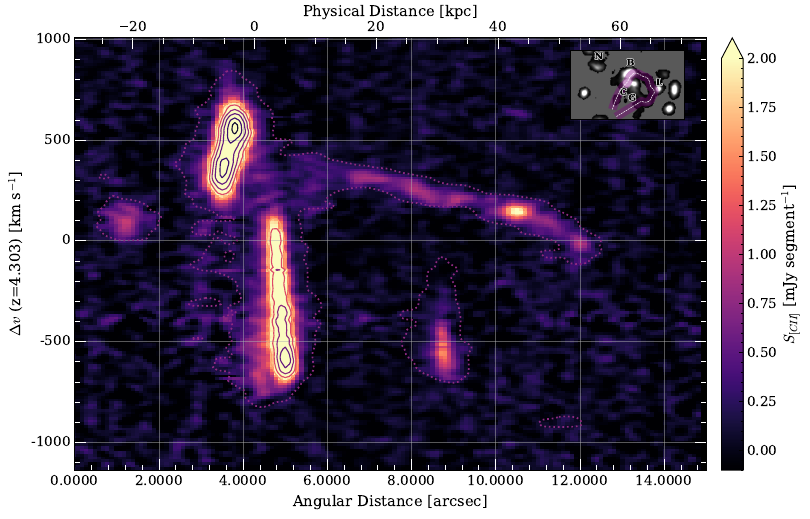}\hspace{-2mm}
\includegraphics[height=0.35\textwidth]{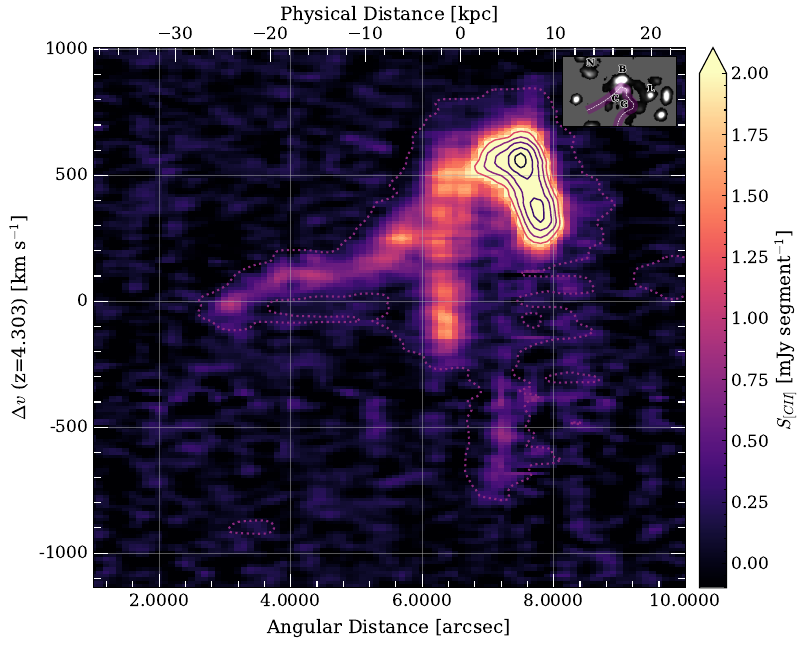}
\caption{The top-center panel shows the same view as in Fig.~\ref{fig:pv} but without including `S5' in the extraction path. The center-left panel shows a path over the counter-streamer system with the SMG `C21' ($-40$ kpc). The center-right PV diagram is for a path that contains follows the streamer from the north of `C', over `S5' (without a cross-over loop) and returns to `C' while only incompletely sampling `G' and `B'. The full PV diagram of `$\nabla$(S3-7)' (with `S5'), starting at `C21' is shown in the bottom-left panel. The same view but for the `$\nabla$(S1-3)' (including `S5') is shown in the bottom-right panel.
\label{fig:additional-pv}}
\end{figure}

Supplementing the position-velocity (PV) diagram in Fig.~\ref{fig:pv}, several more cuts in PV are provided in Fig.~\ref{fig:additional-pv}. 

\newpage
\section{Detection of new [CII] line emitters}\label{appendix:newgalaxies}

Beyond the 23 [C\,{\sc ii}] line emitters cataloged in \citealt{Hill2020} plus UV-bright galaxy `LBG3' \citep{Rotermund2021}, we discovered 11 new protocluster member galaxies in the clipped moment maps of SPT2349$-$56~(south) down to $L_\mathrm{[CII]}=5.2\times10^7$\,$L_\sun$. A detection criterion of $\mathrm{S/N}\gtrsim 4.6$ over $\mathrm{FWHM}_\mathrm{[CII]}>50$\,km\,s$^{-1}$ is required for candidates (see Sec.~\ref{sec:reduction}), in order to show up on our deep \texttt{Clipper} maps. 
By collapsing a $M_0$ map from the $\sigma$-clipped, negative spectral cube, we find that the faintest new [C\,{\sc ii}] candidate is brighter than the most significant negative-$M_0$ pixel (at 71\,mJy\,km\,s$^{-1}$\,beam$^{-1}$) by $\sim1\sigma$, emphasizing a low false-positive source detection rate.

\begin{figure}
\epsscale{1.0}
\center
\includegraphics[width=1\textwidth]{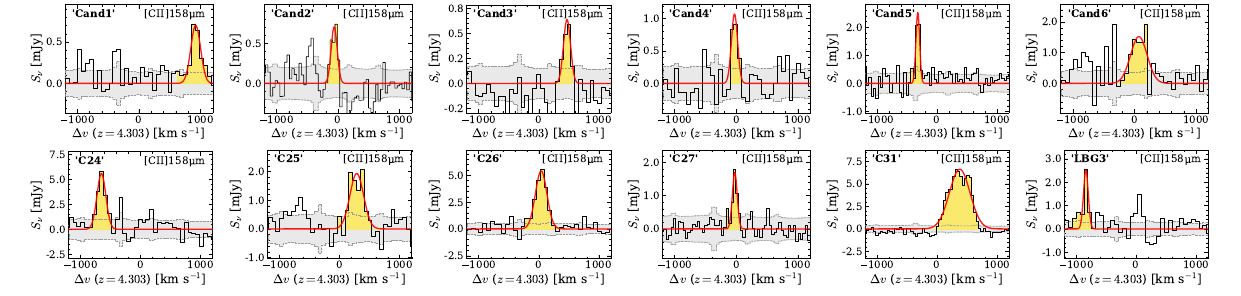}
\caption{[C\,{\sc ii}] spectra of the new galaxies (candidates) extracted from the positions in Tab.~\ref{tab:new_galaxies}. Velocities are centered on $z=4.303$. Red curves indicate the best-fit Gaussian model ($\pm$1$\sigma$ uncertainty on the data; gray fills) employed for flux extraction.
\label{fig:new_galaxies}}
\end{figure}

The spectra of the new galaxies (candidates) are shown in Fig.~\ref{fig:new_galaxies} and the extracted line properties and co-spatial continuum flux detections are provided in Tab.~\ref{tab:new_galaxies}. For sources `C24' through `C31' we keep the nomenclature and employ the same detection method introduced in \citet{Hill2020}. The sources are not ranked by line intensity, since some of the new line emitters are bright in [C\,{\sc ii}] but were previously outside regions with sufficiently high primary beam illumination for detection. Note that `C28', `C29', and `C30' were attributed to the streamer sources `S5',`S4', `S6', respectively, and are thus not duplicated in Tab.~\ref{tab:new_galaxies}. Sources `Cand1' through `Cand6' were not detected with the \citet{Hill2020} method but extracted from the $M_0$ map; they are ranked by the decreasing primary beam illumination.

\begin{deluxetable}{lcccccccccc}[ht!]
\tablenum{5}
\tablecolumns{11}
\tabletypesize{\scriptsize}
\tablecaption{New galaxies in SPT2349$-$56 selected from the clipped [C\,{\sc ii}]158$\mu$m moment maps.
\label{tab:new_galaxies}}
\tablehead{
\colhead{ID} & 
\colhead{R.A. \& Declination} & 
\colhead{$\Omega_\mathrm{gal}$} & 
\colhead{$D_\perp$} & 
\colhead{$S/N$} & 
\colhead{$S_\mathrm{[CII]}\Delta V$} & 
\colhead{$L_\mathrm{[CII]}$} & 
\colhead{$L'_\mathrm{[CII]}$} & 
\colhead{$\Delta v_\mathrm{[CII]}$} & 
\colhead{$\mathrm{FWHM}_\mathrm{[CII]}$} & 
\colhead{$S_{850\mu\mathrm{m}}$} \\
        & 
(J2000) & 
[arcsec$^2$] & 
[kpc] & 
&
$\left[\frac{\mathrm{Jy\,km}}{\mathrm{s}}\right]$ & 
$\left[10^{8} L_{\sun}\right]$ & 
$\left[\frac{10^9\mathrm{K\,km\,pc}^{2}}{\mathrm{s}}\right]$ & 
$\left[\frac{\mathrm{km}}{\mathrm{s}}\right]$ & 
$\left[\frac{\mathrm{km}}{\mathrm{s}}\right]$ & 
[mJy] 
}
\decimalcolnumbers
\startdata
C24  &  23:49:40.52$-$56:38:26.1  &  3.11  &  109.1  &  6.7  &  1.04$\pm$0.22  &  6.18$\pm$1.31  &  2.82$\pm$0.60  &  $-652\pm13$  &  162$\pm$30  &  0.39$\pm$0.29\\
C25  &  23:49:42.58$-$56:38:15.4  &  3.11  &  57.5  &  6.0  &  0.55$\pm$0.12  &  3.30$\pm$0.74  &  1.51$\pm$0.34  &  $284\pm22$  &  261$\pm$50  &  0.21$\pm$0.10\\
C26  &  23:49:41.08$-$56:38:26.7  &  3.11  &  78.8  &  14.0  &  1.31$\pm$0.22  &  7.79$\pm$1.29  &  3.56$\pm$0.59  &  $39\pm8$  &  218$\pm$19  &  1.09$\pm$0.36\\
C27  &  23:49:41.54$-$56:38:21.3  &  1.77  &  52.1  &  5.4  &  0.17$\pm$0.04  &  1.00$\pm$0.24  &  0.46$\pm$0.11  &  $-26\pm9$  &  84$\pm$20  &  $<$0.10\\
C31  &  23:49:42.50$-$56:38:20.3  &  2.42  &  23.4  &  41.7  &  2.86$\pm$0.43  &  17.10$\pm$2.60  &  7.80$\pm$1.18  &  $377\pm5$  &  403$\pm$11  &  1.27$\pm$0.33\\
Cand1  &  23:49:42.65$-$56:38:29.7  &  0.78  &  44.3  &  6.5  &  0.14$\pm$0.03  &  0.84$\pm$0.18  &  0.38$\pm$0.08  &  $921\pm14$  &  176$\pm$34  &  $<$0.06\\
Cand2  &  23:49:42.85$-$56:38:27.2  &  0.78  &  35.3  &  4.8  &  0.09$\pm$0.02  &  0.52$\pm$0.13  &  0.24$\pm$0.06  &  $-63\pm12$  &  110$\pm$28  &  $<$0.06\\
Cand3  &  23:49:41.92$-$56:38:20.9  &  0.79  &  33.5  &  4.3  &  0.09$\pm$0.03  &  0.56$\pm$0.16  &  0.26$\pm$0.07  &  $470\pm16$  &  122$\pm$38  &  $<$0.07\\
Cand4  &  23:49:41.12$-$56:38:23.1  &  0.78  &  73.6  &  4.3  &  0.14$\pm$0.04  &  0.81$\pm$0.22  &  0.37$\pm$0.10  &  $-30\pm13$  &  99$\pm$33  &  $<$0.10\\
Cand5  &  23:49:41.37$-$56:38:16.2  &  0.79  &  78.3  &  4.8  &  0.19$\pm$0.05  &  1.16$\pm$0.30  &  0.53$\pm$0.14  &  $-321\pm7$  &  61$\pm$17  &  0.17$\pm$0.09\\
Cand6  &  23:49:42.66$-$56:38:10.5  &  0.79  &  91.6  &  6.0  &  0.52$\pm$0.12  &  3.08$\pm$0.69  &  1.41$\pm$0.31  &  $68\pm26$  &  311$\pm$60  &  $<$0.16\\
\tableline
LBG3  &  23:49:42.67$-$56:38:28.7  &  3.14 &  38.2   &  7.2  &  0.25$\pm$0.05  &  1.49$\pm$0.30  &  0.68$\pm$0.14  &  $-838\pm8$  &  67$\pm$12  &  $<$0.09\\
\tableline
\enddata
\tablecomments{
See also Tab.~\ref{tab:results} for a description of the columns. $^{(1)}$`LBG3' was previously published in \citet{Rotermund2021}. $^{(3)}$Flux density extraction solid angle $\Omega_\mathrm{gal}$. Circular apertures with a diameter of 1$\arcsec$ are used for galaxy candidates `Cand1'--`Cand6' and 2$\arcsec$ for galaxies `C24'--`C26'. $^{(5)}$The signal-to-noise ratio is obtained from the uncertainty in the flux density amplitude of the Gaussian fit to the spectrum. $^{(11)}$To account for unknown source sizes, the upper limits (3$\sigma$) on the dust continuum flux densities are calculated over an idealized solid angle $\Omega_\mathrm{gal}$ that is equal to the product of the circularized radius of the galaxy solid angles $\Omega_\mathrm{gal}$ times the radius of the circulated beam solid angle $\Omega_\mathrm{bm}$ times $\pi$ (see also Sec.~\ref{sec:continuum}). For detections, prior to adding 15\% calibration uncertainties in quadrature to the variance of $S_{850\mu\mathrm{m}}$ (as it is tabulated here), the significance of the continuum measurements are better than the equivalent 4$\sigma$ upper limits within $\Omega_\mathrm{gal}$ estimated from the background noise.
}
\end{deluxetable}

\bibliographystyle{aasjournal}





\end{document}